\documentclass[aps,prd,twocolumn,notitlepage,nofootinbib,,nobibnotes,superscriptaddress,amsmath,amssymb,preprintnumbers]{revtex4-1}

\usepackage{graphicx} 
\usepackage{dcolumn}
\usepackage{bm}        
\usepackage{amssymb}   
\usepackage{adjustbox}  
\usepackage{amsmath,array}
\usepackage{comment}
\usepackage{natbib}
\usepackage{MnSymbol}
\usepackage[hidelinks]{hyperref}
\usepackage{graphicx}
\usepackage{subfig}
\usepackage{color}
\usepackage{braket}
\usepackage     [utf8]                  {inputenc}
\usepackage     [T1]                    {fontenc}
\usepackage     [english]               {babel}
\usepackage{epigraph}
\usepackage{etoolbox}
\usepackage{ dsfont }
\usepackage{physics}
\usepackage{xcolor}
\usepackage{siunitx}

\makeatletter
\patchcmd{\epigraph}{\@epitext{#1}}{\itshape\@epitext{#1}}{}{}  
\newcommand*\eqsize{%
\@setfontsize\mysize{9.0}{9.0}%
    }
    \usepackage{multirow}
\makeatother

\usepackage{xfrac}
\usepackage{amsmath,amssymb}
\usepackage{esdiff} 
\usepackage{commath} 
\usepackage{bbm} 
\usepackage{braket}
\usepackage{slashed}
\usepackage{caption}
\usepackage{subcaption}

\newcommand{\rmd}{\mathrm{d}}

\newcommand{\rmi}{\mathrm{i}}

\newcommand{\xT}{\mathbf{x}}
\newcommand{\xTp}{\mathbf{x}_1}
\newcommand{\xTm}{\mathbf{x}_2}

\newcommand{\bT}{\mathbf{b}}

\newcommand{\sT}{\mathbf{s}}
\newcommand{\pT}{\mathbf{p}}
\newcommand{\qT}{\mathbf{q}}
\newcommand{\kT}{\mathbf{k}}

\newcommand{\aS}{\alpha_S}

\newcommand{\fm}{\mathrm{fm}}
\newcommand{\GeV}{\;\text{GeV}}

\newcommand{\SNN}{\sqrt{s_{\mathrm{NN}}}}

\newcommand{\Dipper}{\textsc{McDipper}}
\newcommand{\trento}{${\rm T}_{\rm R}$ENTo}

\definecolor{oscar}{RGB}{22, 156, 172}

\begin{document}

\date{\today}

\title{The {\Dipper}: A novel saturation-based 3+1D initial state model for Heavy Ion Collisions}

\author{Oscar Garcia-Montero}
\affiliation{Fakult\"at f\"ur Physik, Universit\"at Bielefeld, D-33615 Bielefeld, Germany}

\author{Hannah Elfner}
\affiliation{GSI Helmholtzzentrum f\"ur Schwerionenforschung, Planckstr. 1, 64291
		Darmstadt, Germany}
\affiliation{Institute for Theoretical Physics, Goethe University,
		Max-von-Laue-Strasse 1, 60438 Frankfurt am Main, Germany}
\affiliation{Frankfurt Institute for Advanced Studies, Ruth-Moufang-Strasse 1, 60438
		Frankfurt am Main, Germany}
\affiliation{Helmholtz Research Academy Hesse for FAIR (HFHF), GSI Helmholtz Center,
		Campus Frankfurt, Max-von-Laue-Straße 12, 60438 Frankfurt am Main, Germany}
\author{Sören Schlichting}
\affiliation{Fakult\"at f\"ur Physik, Universit\"at Bielefeld, D-33615 Bielefeld, Germany}

\begin{abstract} 
We present a new 3D resolved model for the initial state of ultrarelativistic heavy-ion collisions, based on the $k_\perp$-factorized Color Glass Condensate hybrid approach. The {\Dipper} framework responds to the need for a rapidity-resolved initial-state Monte Carlo event generator which can deposit the relevant conserved charges (energy, charge and baryon densities) both in the midrapidity and forward/backward regions of the collision.
This event-by-event generator computes the gluon and (anti-) quark phase-space densities using the IP-Sat model, from where the relevant conserved charges can be computed directly. In the present work we have included the leading order contributions to the light flavor parton densities. As a feature, the model can be systematically improved in the future by adding next-to-leading order calculations (in the CGC hybrid framework), and extended to lower energies by including sub-eikonal corrections the channels included. We present relevant observables, such as the eccentricities and flow decorrelation, as tests of this new approach.
\end{abstract}

\maketitle

\section{Introduction}
It is widely accepted that during heavy ion collisions (HICs) a strongly interacting medium of quarks and gluons is produced, the Quark-Gluon Plasma (QGP)~\cite{Shuryak:2014zxa,Arslandok:2023utm}. The QGP medium expands into the vacuum and after a violent initial pre-equilibrium stage its subsequent space-time evolution can be effectively described by relativistic viscous hydrodynamics. During the expansion of this fluid, spatial anisotropies of the initial state  are converted into momentum space final-state particle multiplicity correlations. These correlations and anisotropies inherited by the final state can be found along both transverse and longitudinal directions, with respect to the collisional axis. 

It is a known result of this area of study that particle correlations in the transverse plane reveal strong azimuthal angle modulation, quantified by the so-called flow harmonics, $v_n$~\cite{Heinz:2013th}. 
However, recently the scrutiny of the longitudinal structure of the $v_n$ and also higher correlations show a non-trivial rapidity dependence \cite{Petersen:2011fp,Nie:2019bgd,CMS:2015xmx,ATLAS:2017rij}. These interesting results call for an improvement of the hydrodynamical models used, which often focus on the midrapidity region, and thus rely on a boost-invariant prescription. The evolution of the medium in a 3+1D fashion has been successfully applied by such hydrodynamical codes as MUSIC~\cite{Schenke:2010nt} and VHLLE~\cite{Karpenko:2013wva}. Once the reader has chosen the model of their preference, what is left is to specify the initial condition for the 3+1D, where they can chose from an array of models, including extended MC-Glauber modeling \cite{Moreland:2014oya,Shen:2020jwv}, deposition via string-hadron transport~\cite{SMASH:2016zqf,Schafer:2021csj}, and color string dynamics \cite{Bozek:2015bna}, amongst others \cite{Ehehalt:1995is, Geiss:1998ki,Bratkovskaya:2011wp}. 

In this work, we present a new framework to compute event-by-event (EbE) longitudinally resolved initial conditions, based on the general ideas of saturation physics. So far initial state models based on saturation physics, such as IP-Glasma \cite{Schenke:2012hg,Schenke:2012wb}, EKRT~\cite{Eskola:1999fc,Niemi:2015qia} or MC-KLN~\cite{Drescher:2006ca} have primarily been designed to describe the initial state around mid-rapidity in 2+1D hydro simulations. Despite various attempts to extend saturation based models to 3+1D ~\cite{Schenke:2016ksl,Schlichting:2020wrv,Schenke:2022mjv,Ipp:2021lwz,Ipp:2022lid}, so far all of these calculations exploit the fact that in the high energy limit the cross-section for gluon production increases rapidly, quickly overtaking that of stopping of valence quarks. However, even in the high energy limit, the so-called fragmentation region in the forward and backward rapidity directions is populated by the stopping of -near- collinear partons containing a high portion of the energy and charge of the colliding nuclei. Evidently, in the forward and backward rapidity regions the net-densities of the conserved electric ($Q$) and baryon ($B$) charges are non-vanishing due to the stopping of valence partons. Hence, to properly describe the 3+1D initial state in heavy-ion collisions, one needs to account not only for the (gluon-dominated) energy deposition at mid-rapidity , but also one for the (valence-quark dominated) energy and charge deposition in the fragmentation region.  Evidently, a proper treatment of the fragmentation region also becomes important, when trying to connect saturation physics based initial state models developed at high-energy, with initial state models at lower energies based e.g. on hadronic transport approaches~\cite{Schafer:2021csj}.

The central objective of this paper is to exploit first principles calculations of quark and gluon scattering processes in high-energy QCD to develop an initial state model for the deposition of energy and the conserved $(u,d,s)$ charges in heavy-ion collisions. Specifically, we will use leading order cross-section calculations in the ${\bf k}_T$-factorized dilute-dense limit of the Color Glass Condensate effective theory of high-energy QCD~\cite{Gribov:1983ivg,Kovchegov:2001sc} to deduce the initial spectra of quark~\cite{Lappi:2012nh,Dumitru:2001jn,Dumitru:2002qt} and gluon production~\cite{Gribov:1983ivg,Kovchegov:2001sc,Gyulassy:1997vt,Blaizot:2010kh,Dumitru:2001ux} and infer the densities of energy and conserved $(u,d,s)$ charges by taking appropriate moments of the spectra. By including event-by-event nucleon position fluctuations as well as realistic parametrizations of the valence parton distributions~\cite{{Buckley:2014ana}} and the dipole scattering amplitude (IPSat)~\cite{Kowalski:2006hc,Kowalski:2003hm}, we develop an efficient Monte Carlo generator -- {\Dipper} for the initial energy and charge density profiles. Within this first study, we further benchmark the model at the hand of multiplicity measurements and provide additional results for the longitudinal structure of initial charge density profiles as well as the longitudinal decorrelation of initial state eccentricities. 

This paper is organised as follows, in sec.~\ref{sec:model} we present an extensive outline of the \Dipper{} overview, where in sec.~\ref{sec:model_part_formulas} we detail the master formulas of the framework, namely single gluon and quark production from the $k_\perp$-factorized CGC formalism, as well as the building blocks for the framework. In sec.~\ref{sec:model_saturation} we introduce two different sets of input models, the GBW and the IP-Sat model, which will be used to benchmark the \Dipper{} framework. In sec.~\ref{sec:EbE_MC} we briefly explain the Monte Carlo sampling used to generate the event-by-event initial conditions.  Subsequently, sec.~\ref{sec:model_structure} contains a description of the code, namely an explanation of  how the aforementioned parts and models are used to produce rapidity resolved EbE HICs initial states. The final subsection in the section corresponds to the fitting procedure to the free-parameter of the framework, the so called $K$-factor. 

In sec.~\ref{sec:anatomy} we give an overview of the generalities of the framework,  a study on the different dependencies on the energy and charge deposition. We explore the thickness, energy and space-time rapidity dependencies, for both saturation models presented in section \ref{sec:model_saturation}. This is performed system-independently, meaning that we analyse charge deposition before the event generation. In sec.~\ref{sec:observables} we explore the EbE observables of the code, where we estimate multiplicity, finding good agreement with data. We also present results for rapidity resolved transverse energy and charge deposition, as well as the long-distance initial correlations, the spatial eccentricities. We discuss also the rapidity correlations which arise with the $x$-dependence of the saturation models, where we present computations for flow-decorrelations. We end by summarizing our findings and presenting the future possible research avenues for this framework in sec.~\ref{sec:conclusions}. Additionally, in Appendix~\ref{app:collinear} we review the collinear limit of the single gluon production formula, valid in the far-fragmentation regions. 



\section{Model Description }
\label{sec:model}

The central idea of our initial state model is to calculate the energy and charge deposition in high-energy heavy-ion collisions within the dilute-dense approximation of the CGC~\cite{Gribov:1983ivg,Kovchegov:2001sc,Kowalski:2006hc,Kowalski:2003hm}. Based on this formalism the single inclusive gluon $(\rmd N_{g}/{\rmd^2\xT \rmd^2\pT \rmd y})$ and (net-) quark distributions $\rmd N_{\bar{q}-q}/{\rmd^2\xT \rmd^2\pT \rmd y}$ as a function of (momentum) rapidity $y$ transverse momentum $\pT$ and transverse position $\xT$ are computed by evaluating the leading order cross-sections for the transverse momentum dependent dipole gluon distributions from the IP-Sat 
model~\cite{Bartels:2002cj,Kowalski:2003hm,Rezaeian:2012ji} and collinear parton distributions (PDFs) from the LHAPDF library~\cite{Buckley:2014ana}. Since at leading order in the high-energy limit the phase-space distributions are proportional to $\delta(y-\eta_{s})$~\cite{Greif:2017bnr}, where $\eta_{s}$ is the space-time rapidity, these distributions also provide the structure of initial state in space-time rapidity upon the straightforward identification of $y=\eta_{s}$. By taking moments of the single inclusive gluon and (net-) quark distributions, we can then directly compute the initial 3+1D profiles of energy ($e$) and  $u,d,s$ charge distributions $(n_{u/d/s})$. By including also fluctuating nucleon positions according to the Monte-Carlo Glauber model~\cite{Miller:2007ri, Alver:2010gr}, our publicly available implementation {\Dipper}~\cite{McDIPPER} thereby provides event-by-event initial conditions for all the conserved charges in the light flavour sector of QCD.

Since our model is based on high-energy QCD cross-section calculations, it has the advantages that a) the number of free parameters in the model is extremely small and b) it can in principle be improved to higher order accuracy as next-to-leading order cross-section calculations become available. Since most of the input information can be extracted alternatively e.g. from Deep Inelastic Scattering (DIS) experiments, the current version of the model only has a single free parameter, namely -- as we will discuss shortly -- a so-called $K$-factor, entering the normalization of the gluon production cross-section.

\begin{figure*}
  \begin{center}
\includegraphics[width=.68\linewidth]{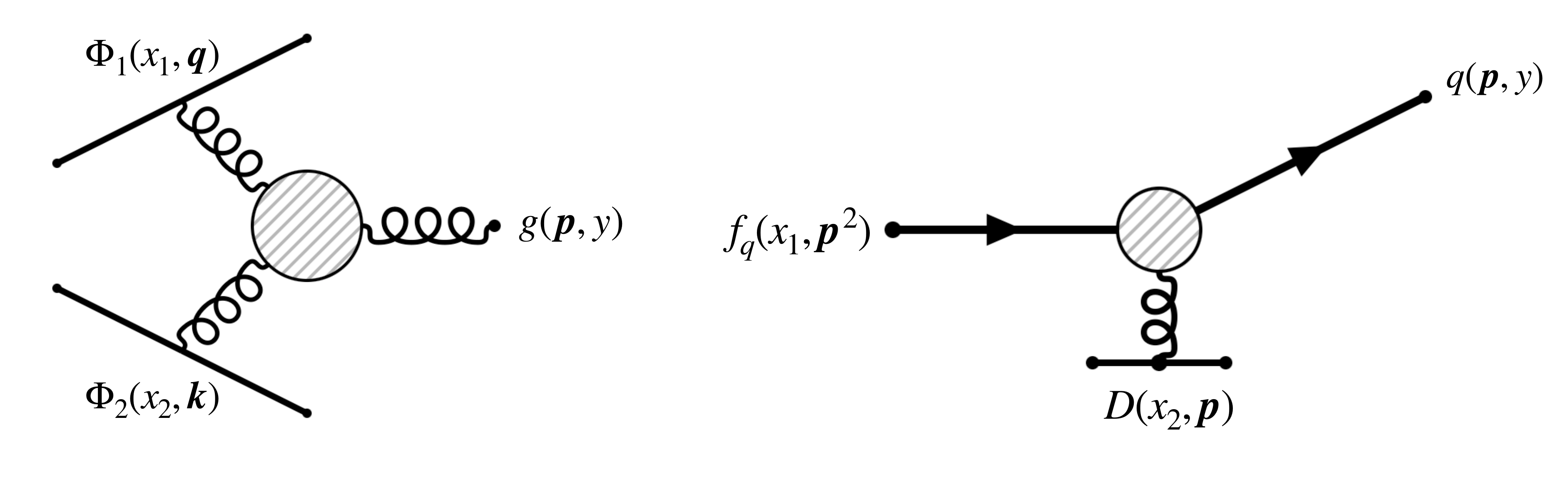}
 \end{center}
  \caption{Diagrammatic representation for the leading order processes included in the {\Dipper} framework. (\emph{Left}) Single gluon production of a gluon with transverse moment $\pT$ and rapidity $y$ from the interaction of two CGCs with gluon distributions $\Phi_{1}(x_1, \qT)$  and $\Phi_{2}(x_2, \kT)$. (\emph{Right}) Deposition of a single quark with transverse momentum $\pT$ an rapidity $y$ from the stopping of a collinear quark from the incoming projectile with momentum fraction $x$ and virtuality $\pT^2$. }
\label{fig:diagrams}
\end{figure*}

\subsection{Single inclusive gluon and quark production}
\label{sec:model_part_formulas}

Now that we have laid out the basic ideas underlying the {\Dipper} we proceed to provide the detailed expressions for the single inclusive gluon production and valence quark stopping cross-sections from the literature along with the description of the calculation of the associated contributions to the energy and charge densities. In order to provide a compact overview the leading order diagrams included in {\Dipper} are depicted in Fig.~\ref{fig:diagrams}.

Based on the $k_\perp$-factorized CGC formalism the single inclusive spectrum of gluons $\frac{dN_{g}}{d^2\xT d^2\pT dy}$ produced in a high-energy collision as a function of transverse momentum $\pT$, momentum rapidity $y$ and transverse position $\xT$, can be expressed as the convolution of the un-integrated gluon distributions $\Phi(x,\xT,\qT)$ of the projectile and target as~\cite{Dumitru:2001ux,Lappi:2017skr}
\begin{equation}
\begin{split}
\frac{dN_{g}}{d^2\xT d^2\pT dy}=  &\frac{g^2\,N_c}{4\pi^5 (N_c^2-1)\,\pT^2}  \int \frac{d^2\qT}{(2\pi)^2}   \frac{d^2\kT}{(2\pi)^2}~\\
&\times~\Phi_1(x_1,\xT,\qT)~\Phi_2(x_2,\xT,\kT)\\
&\times~(2\pi)^2\delta^{(2)}(\qT+\kT-\pT)\\
\end{split}
\label{eq:gluon_prod_w_uGDFs}
\end{equation}
where $x_{1/2}=\frac{\pT}{\sqrt{s_{NN}}} e^{\pm y}$ denotes the light-cone momentum fraction in the projectile and target, $g$ is the strong coupling constant, and $N_c=3$ denotes the number of colors. We note that the above formula, which exhibits $\kT_{\bot}$-factorization,
can be rigorously derived in the case of dilute-dense or dense-dilute treatment of the projectile-target system. However, it has also been demonstrated explicitly that Eq.~\eqref{eq:gluon_prod_w_uGDFs} provides a very good description of the energy deposition even in the case of dense-dense collisions~ \cite{Blaizot:2010kh,Schlichting:2019bvy}, which justify its use in the present context. 

Within the CGC framework of small-x physics, the unintegrated gluon distribution function (uGDF) of the  participant nucleus $i=(1,2)$ can be expressed 
in terms of the correlation function of light-like Wilson lines~\cite{Blaizot:2004wv,Gelis:2001da,Gelis:2001dh}. Specifically,  
\begin{equation}
    \Phi_i(x_i,\xT,\qT) = \frac{4\pi^2 (N^2_c-1)}{g^2 N_c}\,\kT^2\, D_{\rm adj}(x,\xT,\qT)\,
    \label{eq:uGDF}
\end{equation}
where the adjoint dipole $D_{\rm adj}(x,\xT,\qT)$ is defined in momentum space as 
\begin{equation}
D_{\rm adj}(x,\xT,\qT)=\frac{1}{N_c^2-1} \int_{\sT} \tr_{\rm adj}[U^{\rm adj}_{\xT+\sT/2} U^{\rm adj,\dagger}_{\xT-\sT/2}]~e^{i\qT\cdot\sT}\;.
\label{eq:dipole}
\end{equation}
Here $U^{\rm adj}_{\xT+\sT/2}$ and $U^{\rm adj,\dagger}_{\xT-\sT/2}$ denote adjoint light-like Wilson lines at transverse positions $\xT\pm\sT/2$,  such that the transverse momentum scale, $\qT$, is the Fourier transform of the relative size of the dipole, $\sT$, with fixed average position $\xT$. By substituting the above definitions into Eq.~\eqref{eq:gluon_prod_w_uGDFs} one obtains the following expression for single gluon production explicitly in terms of the (adjoint) dipole amplitude, 
\begin{equation}
\begin{split}
\frac{dN_{g}}{d^2\xT d^2\pT dy}=  &\frac{N_c^2-1}{16\pi^4\alpha_s N_c}  \int \frac{d^2\qT}{(2\pi)^2}   \frac{d^2\kT}{(2\pi)^2}~\frac{\qT^2 \kT^2}{\pT^2}\\
&\times~D_{\rm adj}(x_1,\xT,\qT)~D_{\rm adj}(x_2,\xT,\kT)\\
&\times~(2\pi)^2\delta(\qT+\kT-\pT)\\
\end{split}
\label{eq:gluon_prod_dipole}
\end{equation}

While at leading order the single inclusive production of gluons is due to radiative processes, the leading order contribution to single inclusive quark production in the high-energy limit originates from the stopping of collinear quarks inside the projectile/target nuclei due to multi-scattering off the constituents of the other nucleus. By accounting for the stopping of quarks and anti-quarks from both nuclei, the single inclusive distribution $\frac{dN_{q_{f}}}{d^2\xT d^2\pT dy}$
for an (anti-)quark of flavor $f(\bar{f})$ can be expressed as~\cite{Dumitru:2002qt,Dumitru:2005gt}
\begin{equation}
\begin{split}
\frac{dN_{q_{f}}}{d^2\xT d^2\pT dy} &= \frac{x_{1}q^{A}_{f}(x_{1},\pT^2,\xT)~D_{\rm fun}(x_2,\xT,\pT)}{(2\pi)^2} \\
&+ \frac{x_{2}q^{A}_{f}(x_{2},\pT^2,\xT)~D_{\rm fun}(x_1,\xT,\pT)}{(2\pi)^2}\,.
\end{split}
\label{eq:single_quark_density}
\end{equation}
where $q^{A}_{f}(x_{1/2},Q^2,\xT)$ are the collinear quark distributions sinde the two nuclei (see below) and the kinematics of $x_{1,2}$ is the same as for single inclusive gluon production. Intuitively, this expression can be understood as the deflection of a (collinear) quark inside the projectile due to the transverse momentum transfer accumulated in multiple interactions with gluons from the target nucleus. By including all possibilities of multiple scattering, the quark receives a transverse position dependent color rotation expressed in terms of a light-like (fundamental) Wilson line in the amplitude and complex conjugate amplitude. It is then straightforward to show 
that the probability of producing an in-medium quark (non-collinear) as in Eq.~\eqref{eq:single_quark_density}, is just the product of the probabilities of finding a collinear quark of flavor $f$ in the incoming projectile with the probability of a single quark to be deflected by scattering off the target~\cite{Mantysaari:2015uca}, as described by the fundamental dipole distribution
\begin{equation}
    D_{\rm fun}(x,\xT,\qT)=\frac{1}{N_c} \int_{\sT} \tr_{\rm fun}[U_{\xT+\sT/2} U^{\dagger}_{\xT-\sT/2}]~e^{\rm i \qT\cdot\sT} \\ 
\end{equation}


By assuming local Gaussian correlations of color charges inside the nucleus, the fundamental and adjoint dipole gluon distribution in Eq.~\eqref{eq:gluon_prod_dipole} and \eqref{eq:single_quark_density} can be related to each other (in coordinate space) as
\begin{equation}
\label{eq:DFunToDAdj}
D_{\rm adj}(x,\xT,\sT)=[D_{\rm fun}(x,\xT,\sT)]^{C_A/C_F}\;,
\end{equation}
and we will employ this relation throughout this work. With regards to the collinear quark/anti-quark distributions, we simply assume that these can be described in terms of uncorrelated partons, which are distributed in transverse space according to the local density of protons ($p$) and neutrons ($n$). The corresponding distributions are then given by
\begin{equation}
    \begin{split}
    u^A(x,Q^2,\xT)&=u_{p}(x,Q^2)\; T_{p}(\xT) + u_{n}(x,Q^2)\; T_{n}(\xT)\,, \\
     d^A(x,Q^2,\xT)&= d_{p}(x,Q^2)\; T_{p}(\xT) + d_{n}(x,Q^2)\; T_{n}(\xT)\,, \\
      s^A(x,Q^2,\xT)&= T_{p}(\xT)\,s_{p}(x,Q^2) + T_{n}(\xT) \,s_{n}(x,Q^2)\,. \\
      \label{eq:PDFS-pn}
    \end{split}
\end{equation}
for $u,d,s$ flavor quarks/anti-quarks, where by $T_{p/n}(x)=\sum_{i=1}^{A}~t(\xT-\xT_i)\,\delta_{i,p/n}$ we denote the thickness function of protons and neutrons inside the nucleus, where
\begin{equation}
    t(\xT)=\frac{1}{2\pi B_G}\exp\left[-\frac{\xT^2}{2\,B_G}\right]\,,
    \label{eq:thickness}
\end{equation}
where the nucleon size, $B_G=0.156 \fm^{2}$, is determined through fits to the HERA data \cite{Kowalski:2003hm,Rezaeian:2012ji}. The thickness function of an individual nucleon is then normalized as $\int d^2\xT~t(\xT)=1$ such that $\int d^2\xT \,[T_{p}(\xT) +T_{n}(\xT)]=A$ for each nucleus. In this work we will only use a gaussian profile for the individual nucleons.  

Since the parton distributions of the neutron are not well constrained, we will assume isospin symmetry, which means that the neutron PDFs are equivalent to the proton PDFs after applying $u\leftrightarrow d$, i.e. $u_{p/n}=d_{n/p}$, such that Eq.~\eqref{eq:PDFS-pn} simplifies to
\begin{equation}
    \begin{split}
    u^A(x,Q^2,\xT)&=u(x,Q^2)\; T_{p}(\xT) + d(x,Q^2)\; T_{n}(\xT)\,, \\
     d^A(x,Q^2,\xT)&= d(x,Q^2)\; T_{p}(\xT) + u(x,Q^2)\; T_{n}(\xT)\,, \\
      s^A(x,Q^2,\xT)&= \left( T_{p}(\xT) + T_{n}(\xT) \right)\,s(x,Q^2)\,. \\
      \label{eq:PDFS-pn-iso}
    \end{split}
\end{equation}
where we dropped the sub-script $(p)$ in all distributions.

While at first sight, the expression in Eqns.~\eqref{eq:gluon_prod_w_uGDFs} and \eqref{eq:single_quark_density} appear to be rather distinct, it is worth noting
 that in the forward limit, the single gluon production formula yields a result equivalent in form to Eq. \eqref{eq:single_quark_density}. In this limit, the gluon distribution of the projectile is evaluated at large $x$, where the projectile becomes dilute, and the uGDF can be expanded to leading twist~\cite{Blaizot:2004wu,Blaizot:2004wv}. By integrating over the small intrinsic transverse momentum of the projectile gluon (up to the transverse monentum transfer from the target), one the uGDF then reduces to a collinear gluon PDF yielding an expression similar to Eq.~\eqref{eq:single_quark_density} (see Appendix \ref{app:collinear}).  We finally note that -- in the same spirit -- the renormalization scale $Q^2$ of the quark distributions in eq. \eqref{eq:single_quark_density} is set equal to the transverse momentum transfer from the target nucleus ($Q^2=\pT^2$), as this choice ensures that the transverse momentum of the quark remains smaller than the transverse momentum transfer $\pT$ from the target, justifying the collinear factorization treatment. 

Based on the single inclusive spectra of quarks and gluons in Eqns.~\eqref{eq:gluon_prod_w_uGDFs} and \eqref{eq:single_quark_density}, it is then straightforward to compute the local energy density and charge distributions. Specifically, the total energy density is determined as the first moment of the single particle distributions, namely \footnote{We note that in our numerical implementation, we add an infrared regulator by virtue of the replacement $1/\pT^2 \to 1/(\pT^2+m^2)$ to circumvent numerical problems when computing the contribution around an integrable singularity. However, since the integral is UV dominated and infrared finite even in the absence of a regulator, this regulator has no significant effect on the results.}
\begin{equation}
(e\tau)_0 =\int d^2\pT~|\pT|~\left[K_g \frac{dN_{g}}{d^2\xT d^2\pT dy} + \sum_{f,\bar{f}} \frac{dN_{q_f}}{d^2\xT d^2\pT dy}\right]_{y=\eta_s}\;,
\label{eq:EnergyWithKFactor}
\end{equation}
while the conserved charges, $n_i\quad \text{with} i=(u,d,s)$, are computed using the zeroth moment of the distribution, namely
\begin{equation}
(n_f\tau)_0 =\int d^2\pT~\left[\frac{dN_{q_f}}{d^2\xT d^2\pT dy} - \frac{dN_{\bar{q}_f}}{d^2\xT d^2\pT dy}\right]_{y=\eta_s} 
\label{eq:QuarkCharges}
\end{equation}

We note that the above expressions for single inclusive quark and gluon production provide the leading order process in the high-energy limit. In the future, this treatment could be systematically improved by including higher order corrections, such as e.g. the contributions of quark-antiquark pair production by gluon fusion~\cite{Gelis:2003vh,Blaizot:2004wv}. Naturally, a systematic order-by-order improvement of the model also applies to the inclusive gluon production processes, which will include radiative corrections at next-to-leading order~\cite{Iancu:2016vyg,Altinoluk:2018uax,Dumitru:2005gt}.

\subsection{Saturation models}
\label{sec:model_saturation}
Now that we have established the formalism to compute the energy and conserved charge deposition, we continue to discuss the relevant   input from saturation physics. Within this exploratory study, we will consider two different saturation models, namely the GBW Model and the IP-Sat Model, noting that in both cases, the relevant parameters have been determined previously from the analysis of DIS data.  
\subsubsection{GBW Model}
\label{sec:gbw}
We first consider a simple impact parameter dependent generalization of the Golec-Biernat Wusthoff (GBW) model~\cite{Golec-Biernat:1999qor,Kowalski:2003hm,Kowalski:2006hc}, where the fundamental dipole gluon distribution is given by
\begin{eqnarray}
D_{\rm fun}(\xT,\qT)=\frac{4\pi}{Q_F^2(x,\xT)} \exp\left[-\frac{\kT^2}{Q_F^2(x,\xT)}\right]\,.
\label{eq:GBWDip}
\end{eqnarray}

Clearly, the key advantage of this model lies in its simplicity. 
Due to the particularly simple Gaussian form of the dipole gluon distribution, it is possible to obtain analytic expressions for the the dipole gluon distribution both in  transverse position and momentum
space. Similarly, the adjoint dipole distribution also exhibits the same Gaussian shape, with the difference that the saturation scale is replaced by the adjoint saturation scale $Q^2_F\rightarrow Q^2_A$ with $Q_A^2 = (C_A/C_F)\,Q_F^2$. Based on fits to DIS data~\cite{Kowalski:2003hm,Rezaeian:2012ji}, the kinematic dependence of the saturation scale is given phenomenologically by the relation
\begin{equation}
Q_{F}^2(x,\xT)=Q_{p,0}^2\,x^{-\lambda}\,(1-x)^\delta (T_p(\xT)+T_{n}(\xT))\sigma_0
\end{equation}
where $Q_{p,0}^2 = 0.152\,$GeV$^2$, $\lambda=0.215$, $\sigma_0\equiv 2\pi B_G$ denotes the effective transverse area of the nucleon, such that $(T_p(\xT)+T_{n}(\xT))\sigma_0$ essentially counts the density of nucleons per unit transverse area and $\delta=1$ is introduced to regulate the large $x$ behavior.

\subsubsection{IP-Sat Model}
\label{sec:IPSAT}
We will also consider the IP-Sat model \cite{Kowalski:2006hc,Kowalski:2003hm} which provides a more realistic expression for the interaction of a color dipole with a nucleus, as it includes the perturbative tail of large transverse momentum transfers. In the IP-Sat model, the fundamental dipole is given by
\begin{equation}
  D_{\rm fun}(x,\xT,\sT)=\exp\left[-\frac{\pi^2\sT^2}{2\,Nc}\aS(\mu^2)\, xg(x,\mu^2)\, T(\xT)\right]  
  \label{eq:IPSatDip}
\end{equation}
where $\mu^2 = \mu^2_0 + C/\sT^2$, and $g(x,\mu^2)$ is the parton distribution function, which is initialized at a scale $\mu_0$ via the parametrization 
\begin{equation}
xg(x,\mu^2_0)= a_g\,x^{-\lambda_g}\, (1-x)^{5.6}\,.
\end{equation}
This initial condition is evolved along the kinematic range of $\mu^2$ using LO DGLAP evolution. In this work we include the results using the parameter set extracted in ref.~\cite{Rezaeian:2012ji} which fixes the mass of the charm quark mass to $m_c =1.27$ GeV, called in this framework \textit{Set 1}. In the code we also include the necessary files to generate data based on a second parameter set, for which the mass of the charm is fixed to $m_c =1.4$ GeV, or \textit{Set 2}.

Based on the above IP-SAT parametrization, the adjoint dipole can be extracted in coordinate space according to Eq.~\eqref{eq:DFunToDAdj} and the  corresponding momentum space gluon distributions required for the calculation of the energy and charge deposition are then obtained by a numerical Fourier transform with respect to the relative transverse position $\sT$.

\subsection{Event-by-event sampling }
\label{sec:EbE_MC}
Based on the above formalism, we can obtain the average energy and charge deposition profiles for a given configuration of the nucleon positions.  Evidently, it would be possible to include further sub-nucleonic fluctuations e.g. of the spatial and momentum distribution of valence partons, however this is beyond the scope of the present work and we will restrict ourselves to event-by-event fluctuations of the nucleon positions. Our procedure follows common practice, but for the sake of completeness, we will document it below.

Starting point of the event-generation is the sampling of the positions of the individual nucleons in each nucleus from a Wood-Saxon's distribution, for which we include deformation effects for the nuclei which present such properties~\cite{dEnterria:2020dwq}.

Now that we know the positions of each nucleon, two incoming nucleons are set to interact via rejection sampling of the nucleon overlap probability distribution~\cite{McLerran:2015qxa}, 
\begin{equation}
    \frac{\rmd P(b)}{\rmd^2 \bT}=\frac{1-\exp\left(-\sigma_{g}\,T_{NN}(\bT)\right)}{\int\rmd^2\bT \left[1-\exp\left(-\sigma_{g}\,T_{NN}(\bT)\right) \right]}
\end{equation}
where $\sigma_{g}$ represents the total effective partonic cross section at a given collision energy, and $T_{NN}$ stands for the nucleon-nucleon overlap, which can be computed
\begin{equation}
    T_{NN}(\bT)=\int\rmd^2 \sT\,\, t\left(\sT+\frac{\bT}{2}\right)\,t\left(\sT-\frac{\bT}{2}\right)\,.
\end{equation}
Specifically, for the Gaussian nucleon thickness in  Eq.~\eqref{eq:thickness}, the normalization factor of the overlap distribution has the following analytical solution, 
\begin{equation}
\begin{split}
\sigma_{\rm eff}^{\rm inel}(\tilde{\sigma})&=\int \rmd^2 b \left[1-\exp\left(-\sigma_g T_{\mathrm{NN}}(b)\right)\right]\\
&= 4\pi^2\,B_G\left[ \log \left(\frac{\sigma_g}{4\pi B_G}\right)+\Gamma\left(0,\frac{\sigma_g}{4\pi B_G}\right)+\gamma_E\right]
\end{split}
\label{eq:analyticalSigma}
\end{equation}
where $\Gamma(w,z)=\int_z^\infty dt\,t^{w-1}\,e^{-t}$ is the incomplete Gamma function, and $\gamma_E$ is the Euler-Mascheroni constant. By matching $\sigma_{\rm eff,inel}(\sigma_g)=\sigma_{inel}(\SNN)$ to the inleastic nucleon-nucleon cross-section, for which we employ the parametrization in ref.\cite{McLerran:2015qxa}, 

\begin{figure*}
  \includegraphics[scale=0.45]{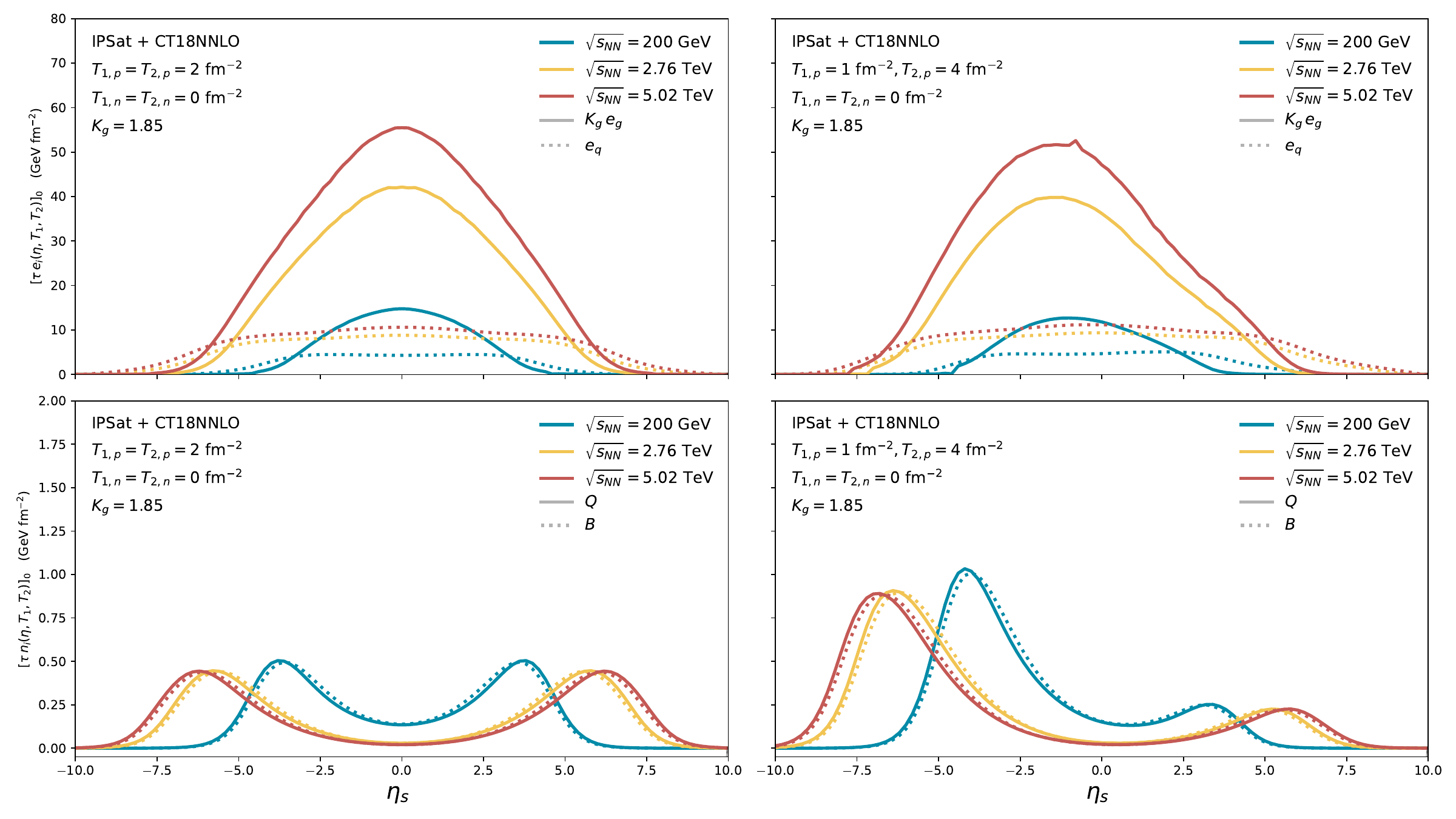}
  \caption{Energy (\emph{upper panels}) and electric and baryon charge deposition (\emph{lower panels}) as a function of space-rapidity, $\eta_s$ for three different collisional energies, , computed using the IP-Sat model with the CT18NNLO PDF set. The results here shown are presented for two different sets of fixed thickness, $T_{1,p}=T_{2,p}=2\,{\rm fm}^{-2}$  (\emph{left panels})  and  $T_{1,p}=1\,{\rm fm}^{-2}$, $T_{2,p}=4\,{\rm fm}^{-2}$ (\emph{right panels}). The neutron density is fixed to zero for both projectiles for this representation, meaning $T_{1,n}=T_{2,n}=0$. 
  }
  \label{fig:energy_and_charges_wrt_eta}
\end{figure*}
\begin{equation}
\begin{split}
\sigma_{\mathrm{inel}}&=
 25.2 + 0.05\log(\SNN) + 0.56\log^2(\SNN) \\
  &+45.2(\SNN)^{-0.9} + 33.8 (\SNN)^{-1.1} \text{mb}
 \end{split}
\end{equation}
we can then determine the $\sigma_{g}$ at each center-of-mass energy\footnote{What it is useful from eq. \eqref{eq:analyticalSigma} is that one can find general dimensionless solutions if one expresses it as 
\begin{equation}
\tilde{\sigma}_{inel}\equiv\frac{\sigma_{inel}}{2\pi\,B_G }=\pi\,\left[ \log \left(\tilde{\sigma}_g\right)+\Gamma\left(0,\tilde{\sigma}_g\right)+\gamma_E\right]
\label{eq:SigmaTildeGen}
\end{equation}
where we have used that $\tilde{\sigma}=\sigma/(2\pi\,B_G)$. This is very convenient, as it will allow the code to have a permanent solution, parametrized in a fit function, regardless of the $B_G$ parameter. The solution of \eqref{eq:SigmaTildeGen} is very well behaved, and can be fitted easily with polynomial function. We use a fifth order polynomial to describe the function accurately (to an error $< 0.05\%$ in the relevant ranges, $50-13000\GeV$ ).}.

Nucleons that undergo at least one inelastic interaction with another nucleon are counted as participants in the collision, and contribute to the overall thickness function $T_{p/n}(\xT)$ of the respective nucleus which determines the energy and charge deposition. Nucleons that do not interact inelastically with any other nucleon are commonly referred to as spectators, and simply disregarded in the calculation of energy and charge deposition.

\subsection{Structure of the Code}
\label{sec:model_structure}
Now that we have described the model, we briefly comment on the numerical implementation {\Dipper}~\cite{McDIPPER}. Clearly, the bottleneck of the numerical implementation of Eqns.~\eqref{eq:EnergyWithKFactor} and~\eqref{eq:QuarkCharges} is the numerical integration required for the calculation of the energy and charge densities for every point in the three dimensional ($\xT$, $\eta_s$) space. However, since for a given center-of-mass energy $\sqrt{s}$ and PDF set the transverse profiles only depend on the space-time rapidity $\eta$ and the nuclear thickness $T_{1/2}$ of the colliding nuclei, the relevant information can be pre-tabulated, allowing for an efficient generation of event-by-event profiles of the conserved charges.

Below we illustrate this at the example of quark production, and refer to the documentation \cite{McDIPPER} of \Dipper{}  for a more concise introduction to the code itself. Single inclusive quark production of a quark with flavor $f$ depends linearly on the nuclear thickness $T_{i}(\xT)$ of the projectile (the nucleus containing the collinear quark) and non-linearly on the thickness $T_{j}(\xT)$ of the target (the nucleus containing small-x gluons), such that e.g. the $u$ charge density per unit rapidity can be expressed as (c.f. Eq.~\eqref{eq:QuarkCharges}
\begin{equation}
\begin{split}
(n_u\tau)_0~=&~~~T_{1,p}(\xT)\left(\,n^{(0)}_{u,12}(T_{2}(\xT))-\,n^{(0)}_{\bar{u},12}(T_{2}(\xT))\right)\\
&+T_{1,n}(\xT)\left(\,n^{(0)}_{d,12}(T_{2}(\xT))-\,n^{(0)}_{\bar{d},12}(T_{2}(\xT))\right)\\
&+T_{2,p}(\xT)\left(\,n^{(0)}_{u,21}(T_{1}(\xT))-\,n^{(0)}_{\bar{u},21}(T_{1}(\xT))\right)\\
&+T_{2,n}(\xT)\left(\,n^{(0)}_{d,21}(T_{1}(\xT))-\,n^{(0)}_{\bar{d},21}(T_{1}(\xT))\right)
\end{split}
\label{eq:nu_quark_decomposition}
\end{equation}
where $T_{i/j,p/n}(\xT)$ is the density of proton/neutrons in nucleus $i/j$ and 
\begin{equation}
n^{(k)}_{f,ij}(T_{j}(\xT)) = \int \rmd^2 \pT \,|\pT|^{k}\, \frac{x_{i}q_{f}(x_{i},\pT^2)~D_{\rm fun}(x_j,T_{j}(\xT),\pT)}{(2\pi)^2} 
\label{eq:quark_decomposition}
\end{equation}
are simple moments of the single-inclusive quark/anti-quark distribution, which depend on the transverse position $\xT$ only through the nuclear thickness $T_{j}(\xT)$. By pre-tabulating the expressions in Eq.~\eqref{eq:quark_decomposition} for a given center-of-mass energy $\sqrt{s}$ as a function of rapidity $\eta$ and thickness $T_{j}(\xT)$, it is then possible to directly evaluate Eq.~\eqref{eq:nu_quark_decomposition} without the need for any additional numerical integration.

Similarly, one can also calculate and pre-tabulate the deposited energy per unit rapidity $(e\tau)_{0}(\xT)$, as a three dimensional function of the space-time rapidity $\eta$ and the nuclear thickness $T_{1}(\xT),T_{2}(\xT)$ of the two nuclei for a given center-of-mass energy $\sqrt{s}$. Even though the calculation of these tables typically requires 2-3 CPU hours\footnote{The running time depends on the collisional energy. Lower energies require relatively shorter computation times, since the more rapidity cells are kinematically restricted or prohibited, which means that the $x$ values in the integrand are outside of the allowed ranges, and therefore less points in the grid need to be accessed.} -for a 3D grid with 101 points in all $T_1,T_2$ and $\eta_s$ directions-, the generation of an individual event is then very efficient, and only requires 20-25 CPU seconds, when computed on a grid with 101 points in rapidity and 256x256 points in the transverse plane. 

\begin{figure*}
  \includegraphics[scale=0.48]{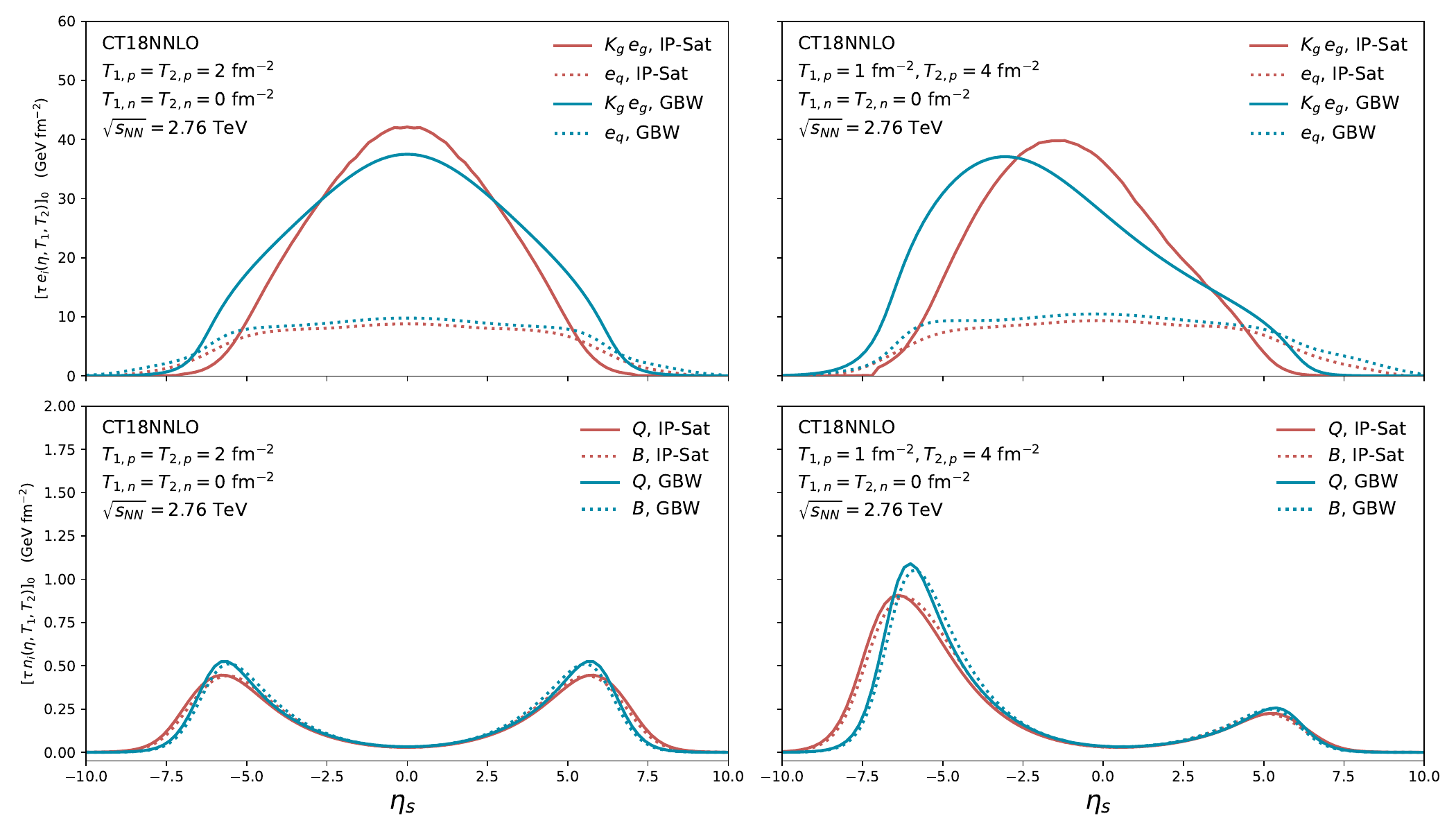}
  \caption{Energy (\emph{upper panels}) and electric and baryon charge deposition (\emph{lower panels}) as a function of space-rapidity, $\eta_s$, for $\sqrt{s_{\rm NN}}=5.02$~TeV, computed using both the IP-Sat and GBW model with the CT18NNLO PDF set. The results here shown are presented for two different sets of fixed thickness, $T_{1,p}=T_{2,p}=2\,{\rm fm}^{-2}$  (\emph{left panels})  and  $T_{1,p}=1\,{\rm fm}^{-2}$, $T_{2,p}=4\,{\rm fm}^{-2}$ (\emph{right panels}). The neutron density is fixed to zero for both projectiles for this representation, meaning $T_{1,n}=T_{2,n}=0$. }
  \label{fig:anatomyGBWvsIPSat}
\end{figure*}

\subsection{Determination of the $K$-factor}
We already explained above, that the there is only one free parameter in the model $K_g$, which enters the calculation of the energy deposition in Eq.~\eqref{eq:EnergyWithKFactor} and supposedly accounts for higher-order corrections in the leading order expression to single inclusive gluon production~in Eq.~\eqref{eq:gluon_prod_w_uGDFs}. 
We note that the introduction of such a parameter is a well known procedure, and has been employed in many previous works using the hybrid formalism \cite{Ma:2017rsu,Ma:2018bax,Benic:2018hvb}. In this work we will determine $K_g$ by matching our results to the experimentally measured values of transverse energy per unit rapidity at mid-rapidity in min.bias p+p and p+Pb collisions, i.e.
 
\begin{equation}
    \left\langle\frac{\rmd E_\perp}{\rmd y}\right\rangle_{\rm exp} =\left\langle\frac{\rmd E_{q,\perp}}{\rmd y}\right\rangle +K_g \,\left\langle\frac{\rmd E_{g,\perp}}{\rmd y}\right\rangle\,,
    \label{eq:matching}
\end{equation}
 Experimental values for the transverse energy per unit rapidity are determined from the ALICE data as in ref.  \cite{ALICE:2022imr},

\begin{equation}
\begin{split}
    \left\langle\frac{\rmd E_\perp}{\rmd y}\right\rangle&= \langle m_
    \perp\rangle \frac{1}{f_{\rm tot}} \frac{\rmd N_{\rm ch}}{\rmd y}\\
    &\approx \frac{ \langle m\rangle }{f_{\rm tot}}\sqrt{1+a^2} J^{-1}(a,\eta)\frac{\rmd N_{\rm ch}}{\rmd \eta}
\end{split}
\end{equation}
where $\langle m_\perp\rangle$ and $\langle m\rangle=0.215\pm 0.001$ are the average transverse mass and rest mass of all measured particles, respectively. The prefactor $f_{\rm tot}=0.55\pm 0.01$ stands for the ratio of charged particles to the total number produced. The parameter $a=\langle p_\perp\rangle/\langle m\rangle$ is used to approximate $\langle m_\perp \rangle$. Additionally, we need the Jacobian $J(a,\eta)= \sqrt{1+1/(a\cosh{\eta})^2}$ to convert the pseudorapidity dependent measured results into rapidity. 

The transverse energy deposited is then matched to the mid-rapidity value for p-p and p-Pb minimum bias collisions, using the relation \eqref{eq:matching}. The experimental errors are propagated accordingly to the $K_g$ factor. Using this prefactor fitting scheme, we obtain $K_{g,\rm GBW}=1.25$ and $K_{g,\rm IP-Sat}=1.85$. 

\section{General features of the model}
\label{sec:anatomy}
Before we proceed to present results for observables relating to the transverse and longitudinal structure of our initial state model, we find it important to further illustrate some of its general features.

We first showcase the rapidity dependence of energy and charge deposition in Fig.~\ref{fig:energy_and_charges_wrt_eta}, where the top panels show the initial energy $(e\tau)_0$ per unit rapidity, carried by quark $(q)$ and gluon $(g)$ degrees of freedom, while the bottom panels show the deposited amount of baryon $(B)$ and electric $(Q)$ charge $(\tau n)_0$ per unit rapidity. Different columns show the space-time rapidity dependence of the deposition profile, for different configurations of the nuclear thickness in the colliding nuclei, for a symmetric configuration $T_{1,p}=T_{2,p}=2 fm^{-2}$ (left) and an asymmetric configuration $T_{1,p}=1 fm^{-2}\;,~T_{2,p}=4 fm^{-2}$ (right). Different colored curves in each panel correspond to two different center of mass energies, $\sqrt{s}_{NN}=200 {\rm GeV}$ (blue) and $\sqrt{s}_{NN}=2.76 {\rm TeV}$ (red). Before we comment on the detailed features, it is important to emphasize, that all the dependencies on space-time rapidity, center-of-mass energy and nuclear thickness are entirely determined by the $x$ and $T$ dependence of the uGDS and PDFs, and if not stated otherwise we employ the IP-Sat and the CT18NNLO PDF sets for the following figures.

When considering symmetric thickness profiles, shown in the left column of Fig.~\ref{fig:energy_and_charges_wrt_eta}, we find that the energy deposition at mid-rapidity is dominated by the gluonic contribution $e_{g}$. Nevertheless both at RHIC and LHC energies the quark contribution to the energy deposition $e_{q}$ is non-negligible. By moving out to larger rapidities, the relative contribution of quarks to the energy deposition increases; at the same time valence charge is deposited at large rapidities resulting in pronounced peaks in the net baryon (B) and net electric charge (Q) distribution.  By comparing the results for different center of mass energies, one finds that the gluon dominance at mid-rapidity increases with increasing center of mass energy, which results in a rapid rise of the overall energy deposition. Conversely, the overall amount of net-charge deposition exhibits a much weaker dependence on the center of mass energy, as the peaks mainly shift out to larger rapidity for increasing center of mass energy.

 \begin{figure*}
  \includegraphics[scale=0.42]{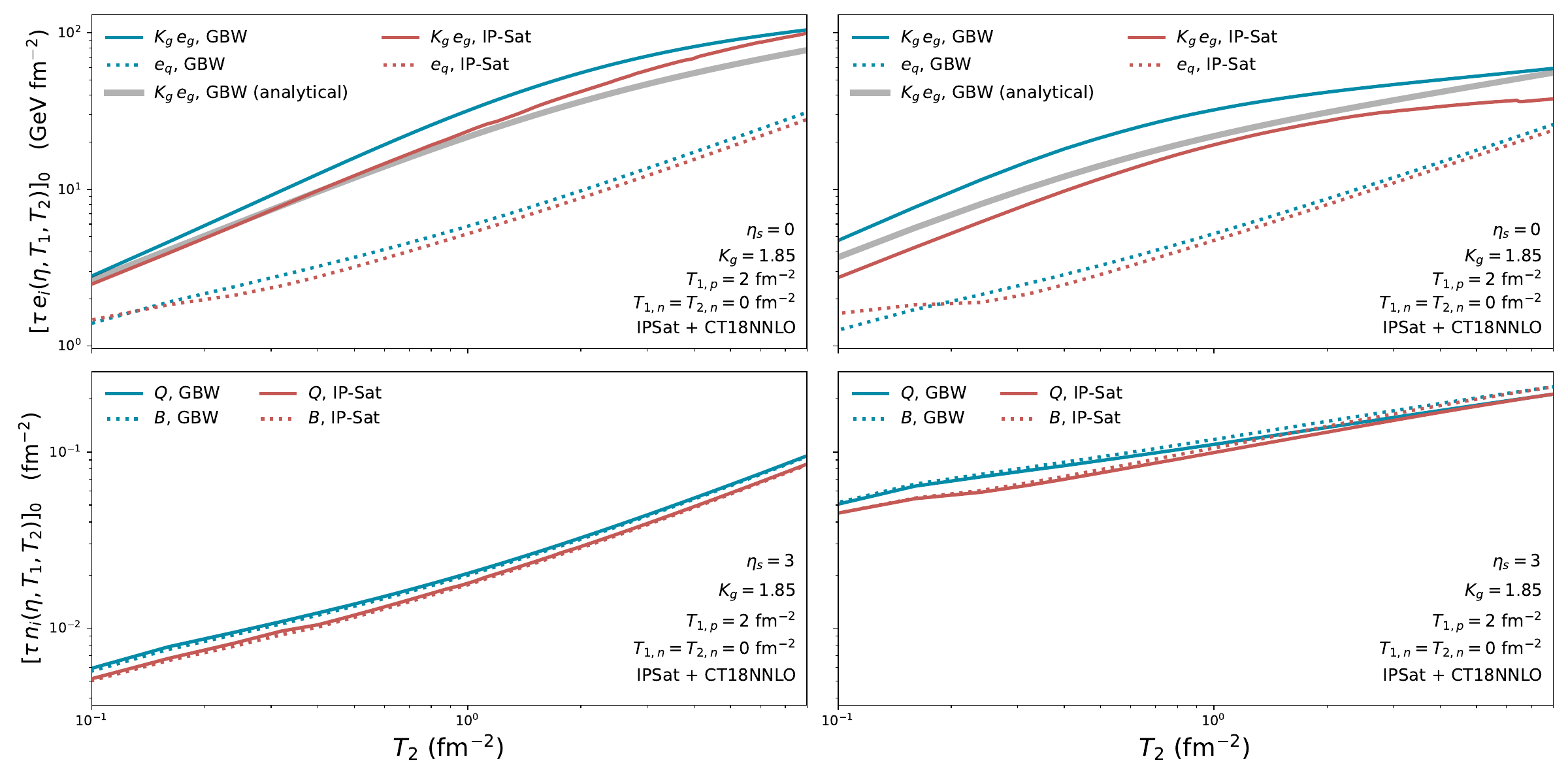}
  \caption{ Energy (\emph{upper panels}) and electric and baryon charge deposition (\emph{lower panels}) as a function of the proton thicknes, $T_{2,p}$, for fixed $T_{1,p}=2\,{\rm fm}^{-2}$ and collisional energy $\sqrt{s_{\rm NN}}=5.02$~TeV, computed using both the IP-Sat and GBW model with the CT18NNLO PDF set. The results here shown are presented for two different sets of fixed space-time rapidity, $\eta_s=0$  (\emph{left panels})  and  $\eta_s=3$ (\emph{right panels}).  }
  \label{fig:Anatomy-ThicknessDep}
\end{figure*}
By looking at the results in the right column of Fig.~\ref{fig:energy_and_charges_wrt_eta},  one finds that  the energy and charge deposition for asymmetric thickness profiles, features highly asymmetric rapidity distributions. Clearly, the peaks in the energy and charge deposition move away from mid rapidity, indicating that geometric fluctutations in the transverse plane can induce non-trivial modulations of the  rapidity profiles on an event-by-event basis.
While the energy deposition due to quarks only shows a relatively weak forward-backward asymmetry, this effect is much more pronounced for the energy deposition of gluons and the charge deposition due to valence quark stopping.

By comparing the results for the IP-Sat and GBW model, as depicted examplarily in Fig.~\ref{fig:anatomyGBWvsIPSat}  for a center of mass energy $\sqrt{s_{NN}}=2.76$ TeV, we can gauge the sensitivity to the underlying UGDs. While the total energy and charge deposition turns out to be rather similar between the two saturation models, moderate differences emerge in the rapidity dependence of the energy and charge deposition, which we attribute to the rather different $\pT$ dependence of the UGDs in the IP-Sat and GBW model.

Next we investigate the dependence of the energy and charge deposition on the nuclear thickness, as shown in Fig.~\ref{fig:Anatomy-ThicknessDep}, where we fix the nuclear thickness $T_{1}^{p}=2 {\rm fm}^{-2}$ of protons of one of the nuclei and vary the nuclear thickness of protons $T_{2}^{p}$ of the other nucleus. Neutron thicknesses of both nuclei $T_{n}^{1/2}$ are set to zero in this comparison. Irrespective of the underlying saturation model (GBW or IP-Sat), the $T_{1}$ and $T_{2}$ dependence energy and charge deposition is governed by the $T$ dependence of the valance parton flux and the saturation scales of both nuclei. Interestingly, the quark and gluon contributions to the energy deposition exhibit a rather different $T$ dependence. In the case of the gluon contribution $e_{g}$, one observes a rapid rise for $Q_{s}^{(2)}<Q_{s}^{(1)}$ followed by a saturation of growth for  $Q_{s}^{(2)}>Q_{s}^{(1)}$. Conversely, the energy deposition due to quark stopping $e_{q}$ exhibits a steady rise across the entire range of $T$ values depicted in Fig.~\ref{fig:Anatomy-ThicknessDep}.
It is particularly important to point out, that due to the asymmetry in the saturation scales, the energy deposition profiles at different rapidities and center of mass energies, are different both in regards to the  $T_{1}$ and $T_{2}$ dependence as well as in the overall normalization. Notably, this behavior can be intuitively understood in the GBW model, if the $x$ dependence of the UGDs in Eq.~\eqref{eq:gluon_prod_w_uGDFs} is approximated by self-consistently evaluating $x_{1/2}=Q_s^{1/2}(x)/\sqrt{s_{NN}}e^{\pm y}$ instead of accounting for full $\pT$ dependence of the UGDs. Since the $\pT$ dependence of $x$ is ignored in this approximation, one can integrate analytically as in ref.~\cite{Borghini:2022iym} to obtain an expression for the energy density. Such an exression will depend on the saturation scales parametrically as follows 

\begin{figure}
  \includegraphics[scale=0.50]{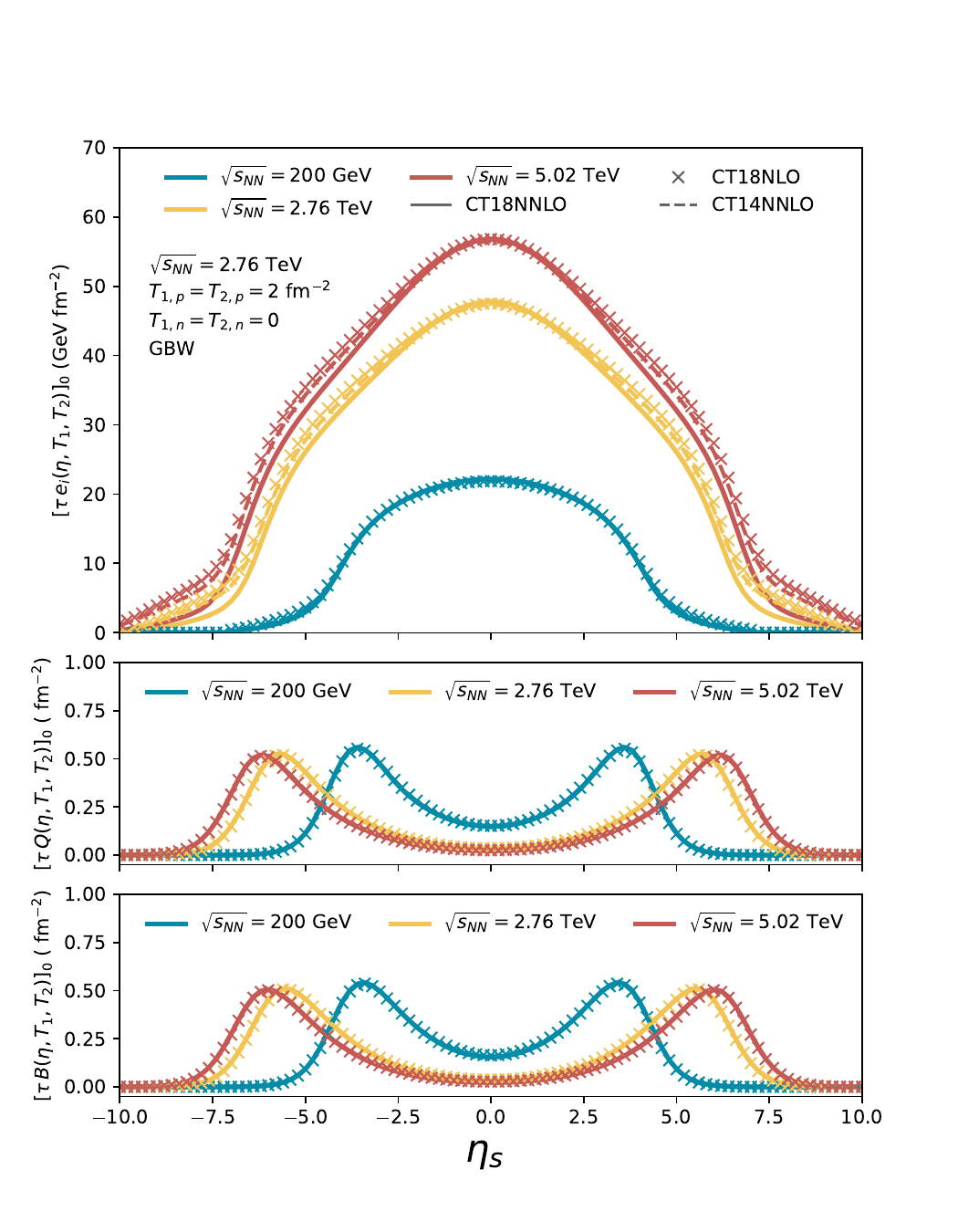}
  \caption{Comparison of the energy density (\emph{top}), electric charge density $Q$ (\emph{middle}) and baryon charge density $B$ (\emph{bottom}) deposition when using different PDFs. Here are pictured the conserved charges when using the CT18NLO, CT18NLO and CT14NNLO sets. As the reader can easily see, while there are significant changes in the quark energy density, $Q$ and $B$ do not change, just as expected.}
  \label{fig:diff_PDF_energy}
\end{figure}

\begin{eqnarray}
    e_{g} \propto \frac{Q_{s,1}^{2}Q_{s,2}^{2}\big(Q_{s,1}^{4}+\frac{7}{2} Q_{s,1}^{2} Q_{s,1}^{2} + Q_{s,2}^{4}\big) }{(Q_{s,1}^{2} +Q_{s,2}^{2})^{5/2}}\,,
    \label{eq:param_GBW}
\end{eqnarray}
where according in the small-$x$ limit, $x_{1/2}=Q_s^{1/2}(x)/\sqrt{s_{NN}}e^{\pm y}$ can be self-consistently solved to fin the analytical expression,  
\begin{eqnarray}
    Q_{s,1/2}^{2} \propto T_{1/2}^{\frac{2}{2+\lambda}} e^{\mp \frac{2\lambda }{2+\lambda}\eta} \Big(\sqrt{{s}_{NN}}\Big)^{\frac{2\lambda}{2+\lambda}}. 
\end{eqnarray}
We depict this result, Eq.~\ref{eq:param_GBW}, as a gray line in Fig.~\ref{fig:Anatomy-ThicknessDep}. Even though such a simple parametrization can approximately describe the behavior over a large range of nuclear thickness, it fails to describe important details that emerge from the non-trivial $x$-dependence of the underlying parton distributions.

Before we turn to phenomenological results, we finally investigate the sensitivity of the energy and charge deposition to the underlying PDF set. We provide a compact summary of our results in Fig.~\ref{fig:diff_PDF_energy}, where we compare the rapidity dependence of energy and charge deposition for the CT18NNLO,CT18NLO and CT14NNLO PDF sets. Since the large $x$ structure of the valence quark PDFs is well constrained, the charge deposition profiles are virtually unchanged. Slight differences emerge for the energy deposition, that probes the sea quark PDFs, in particular at larger rapidities where the small $x$ is explored. Nevertheless, the profiles are remarkably stable with regards to the PDF uncertainties.

\begin{figure}[t]
  \includegraphics[scale=0.55]{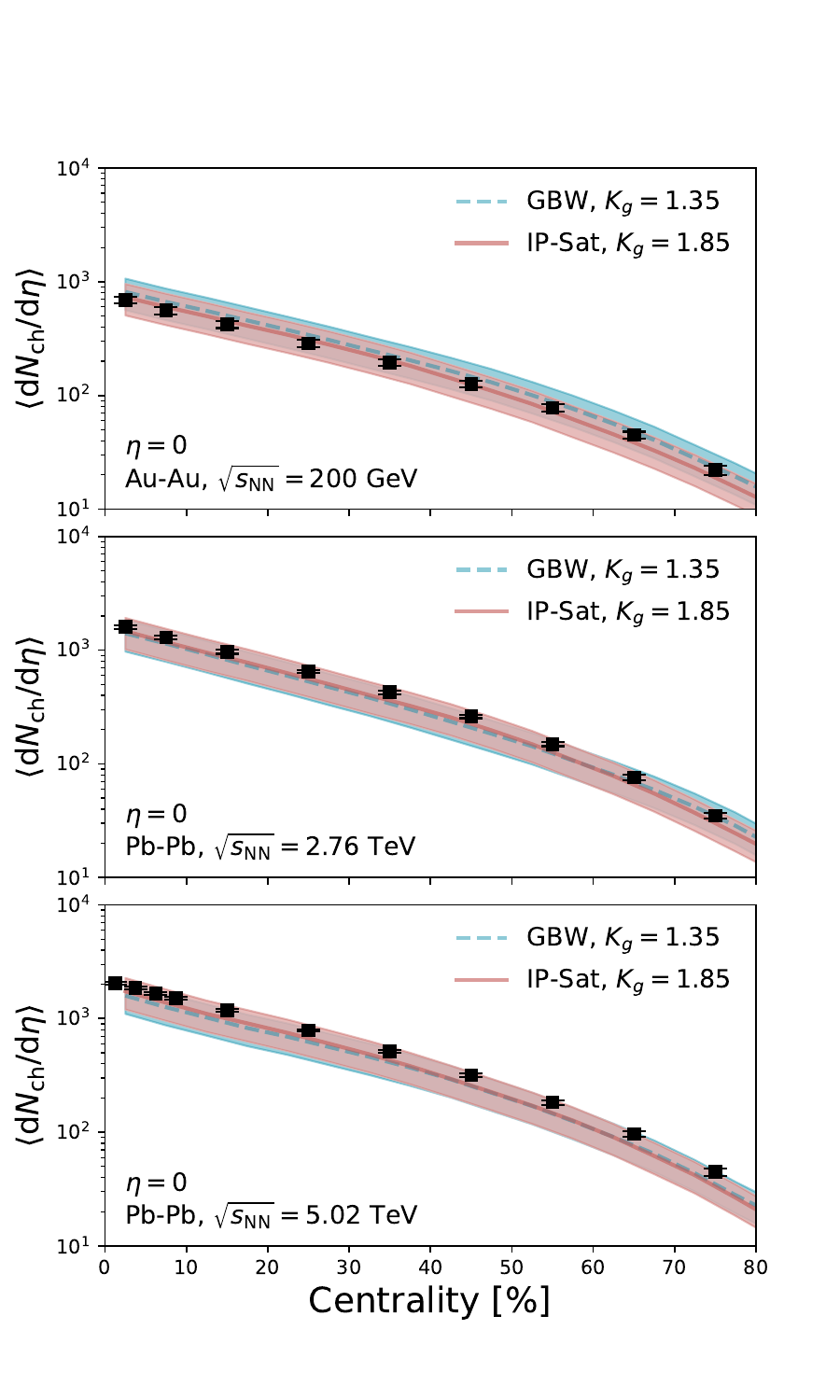}
  \caption{Charged particle yield as a function of centrality for $\sqrt{s_{\rm NN}} =200$ GeV (\emph{top}), $\sqrt{s_{\rm NN}} =2.76$ GeV (\emph{center}) and $\sqrt{s_{\rm NN}} =5.02$ GeV (\emph{Bottom}). Experimental data shown in the figure for Au-Au at  $\sqrt{s_{\rm NN}} =200$~GeV is from  ref.~\cite{STAR:2008med}, Pb-Pb collisions at $\sqrt{s_{\rm NN}} =2.76$~TeV from ref.~\cite{ALICE:2013jfw,ALICE:2010mlf}, and $\sqrt{s_{\rm NN}} =5.02$~TeV  from ref.~\cite{ALICE:2016fbt,ALICE:2015juo}. The bands represent the uncertainties in $\eta/s$ and $C_{\infty}$ for the estimates here presented, where they correspond to variations of $C_{\infty} = 0.8-1.15$ and $\eta/s =0.08-0.24$. }
  \label{fig:dNchdeta_all_energies}
\end{figure}

\section{Observables}
\label{sec:observables}
Now that we have established the basic features of our initial state model, we continue to constrast the model with experimental results at RHIC and LHC. Clearly, the results presented below are only meant to provide a general idea of the behavior of the {\Dipper} model as we anticipate more detailed phenomenological studies to be performed based on full 3+1D Initial State + Hydrodynamics + Hadronic Cascade simulations in the future. 
\begin{figure*}
  \includegraphics[scale=0.45]{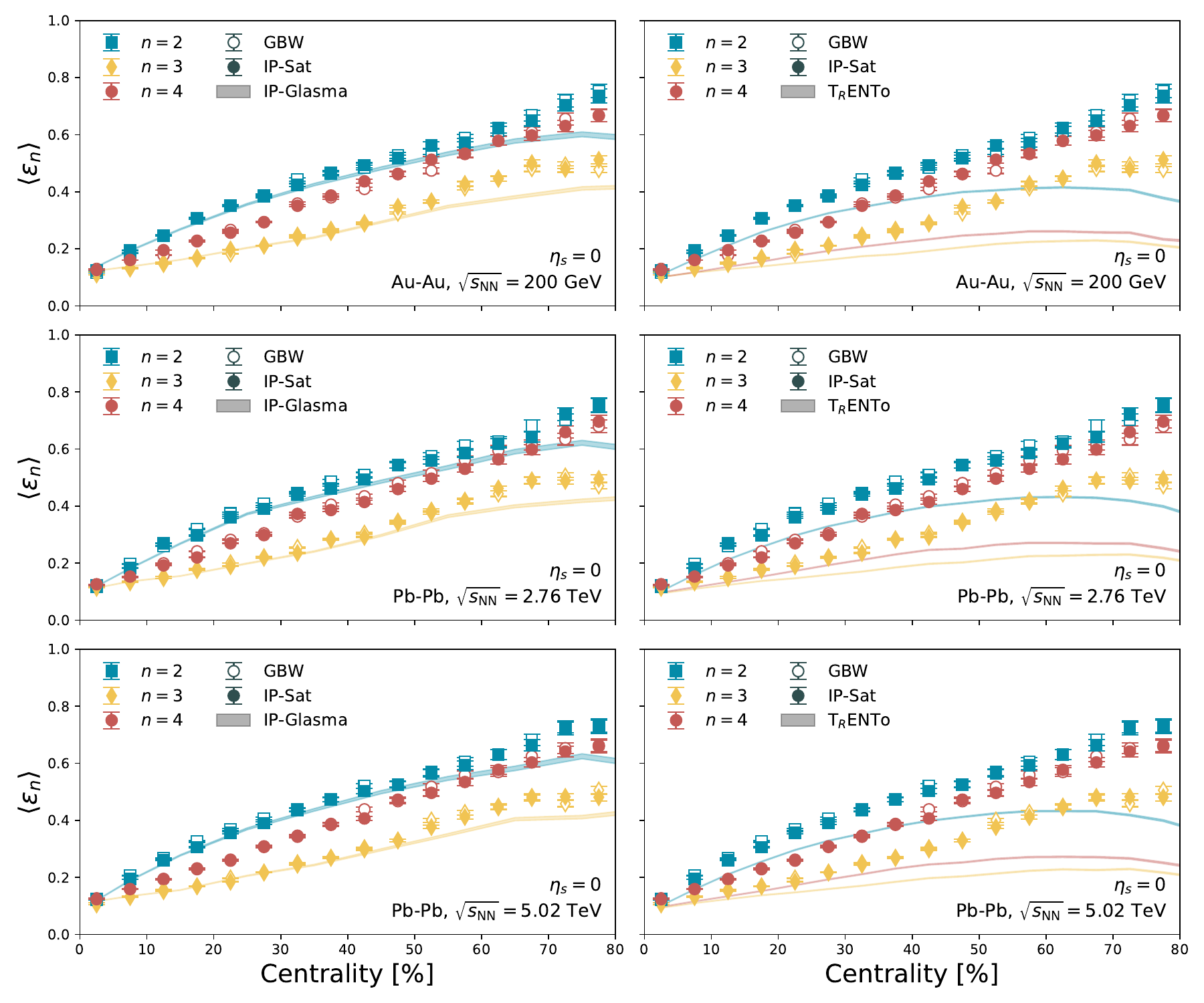}
  \caption{
  	Eccentricities in the IP-Sat (full symbols) and GBW (empty symbols) models as a function of centrality, $\eta_s$. We show here the $n=2,3,4$ harmonics (blue squares, red circles and yellow diamonds, repectively) for Pb-Pb collisions at $2.76$~TeV at midrapidity $\eta_s =0$. Included is a comparison to \trento~\cite{Moreland:2014oya,Borghini:2022iym} and IP-Glasma~\cite{Schenke:2020mbo} models.  }
  \label{fig:ecc_centr}
\end{figure*}

Before we explore the full 3D structure of the model it is important to cross-check that observables at mid-rapidity can be reasonably well described within we model. We first investigate the centrality dependence of the charged particle multiplicity $\rmd N_{\rm ch}/\rmd y$, which following \cite{Giacalone:2019ldn} can be estimated directly from the initial energy density profile as
\begin{equation}
\begin{split}
    \left\langle \frac{\rmd N_{\rm ch}}{\rmd y}\right\rangle =& \frac{4}{3}\frac{N_{\rm ch}}{S} \, C_\infty^{3/4}\,\left(4\pi\,\frac{\eta}{s}\right)^{1/3}\left(\frac{\pi^2}{30}\nu_{\rm eff}\right)^{1/3}\\
    &\times\int \rmd^2\xT \left[\tau e(y,\xT)\right]^{2/3}_0
\end{split}
\end{equation}
where $\nu_{eff}=40$, $S/N_{\rm ch}=7.5$. We present the results in Fig.~\ref{fig:dNchdeta_all_energies}, where we compare the centrality dependence of $\rmd N_{\rm ch}/\rmd y$ to experimental measurement at RHIC \cite{STAR:2008med} and LHC energies~\cite{ALICE:2016fbt,ALICE:2015juo,ALICE:2013jfw,ALICE:2010mlf}. Bands signify the uncertainties in $\eta/s$ and $C_{\infty}$ in these estimates and correspond to variations of $C_{\infty} = 0.8-1.15$ and $\eta/s =0.08-0.24$. We find that within these uncertainties, the model describes the centrality dependence of the data rather well; in particular also the energy dependence of the charged particle multiplicity is well described within the \Dipper.

Next we turn to behaviour of the event-by-event eccentricities at mid-rapidity, which provide knowledge about collective flow and geometric fluctuations. We present our results for the cumulants $\varepsilon_{2}\{2\},\varepsilon_{3}\{2\}$ and $\varepsilon_{4}\{2\}$ in Fig.~\ref{fig:ecc_centr}, where we plot $\varepsilon_{n}\{2\}$ as a function of centrality. The eccentricities are computed using the definition 
\begin{equation}
\varepsilon_n(y) = \frac{\int \rmd^2 r_\perp |\mathbf{r}_\perp|^n\, e^{\rmi n (\phi -\Phi_n)} \left[e(y,\mathbf{r}_\perp)\tau\right]_0 }{\int \rmd^2 r_\perp  \left[e(y,\mathbf{r}_\perp)\tau\right]_0 }
\end{equation}with  $\varepsilon_n\{2\}= \sqrt{\left\langle|\varepsilon_n|^2\right\rangle}$, see ref. \cite{Schenke:2022mjv}.
Different panels show the results for $Au+Au$ collisions at RHIC (top) and Pb+Pb collisions at LHC (center,bottom) and we also compare the results of the {\Dipper} to calculations in the \trento~\cite{Moreland:2014oya,Borghini:2022iym}\footnote{We use parameters extracted from a Bayesian analysis from ref. \cite{Liyanage:2023nds}, namely  $p=0.038$, $w=0.985$ in the {\trento} parametrization. For the parameter $N$, we used $N=6.39$ for Au-Au at $
	\sqrt{s_{\rm NN}}= 200$ GeV, 
	$N=20.013$ for Pb-Pb at $
	\sqrt{s_{\rm NN}}= 2760$ GeV and $N=25.96$ for Pb-Pb at $
\sqrt{s_{\rm NN}}= 5020$ GeV.} and IP-Glasma~\cite{Schenke:2020mbo} models. 
We find excellent agreement with the IP-Glasma model for central collisions $< 50\%$ centrality and small differences for larger centralities. We find that, in comparison to the {\trento} model, the eccentricities obtained in our model are 
rather large, as has already been observed in earlier comparisions of the MC-Glauber and MC-KLN models, which are similar in spirit. Given the fact that the two rather different models TrENTO and IP-Glasma can both describe the experimental results rather well, it seems likely  -- with suitable adjustment of hydrodynamic evolution parameters -- experimental results for collective flow can also be described by the \Dipper.

\begin{figure*}
  \begin{center}
\includegraphics[width=.52\linewidth]{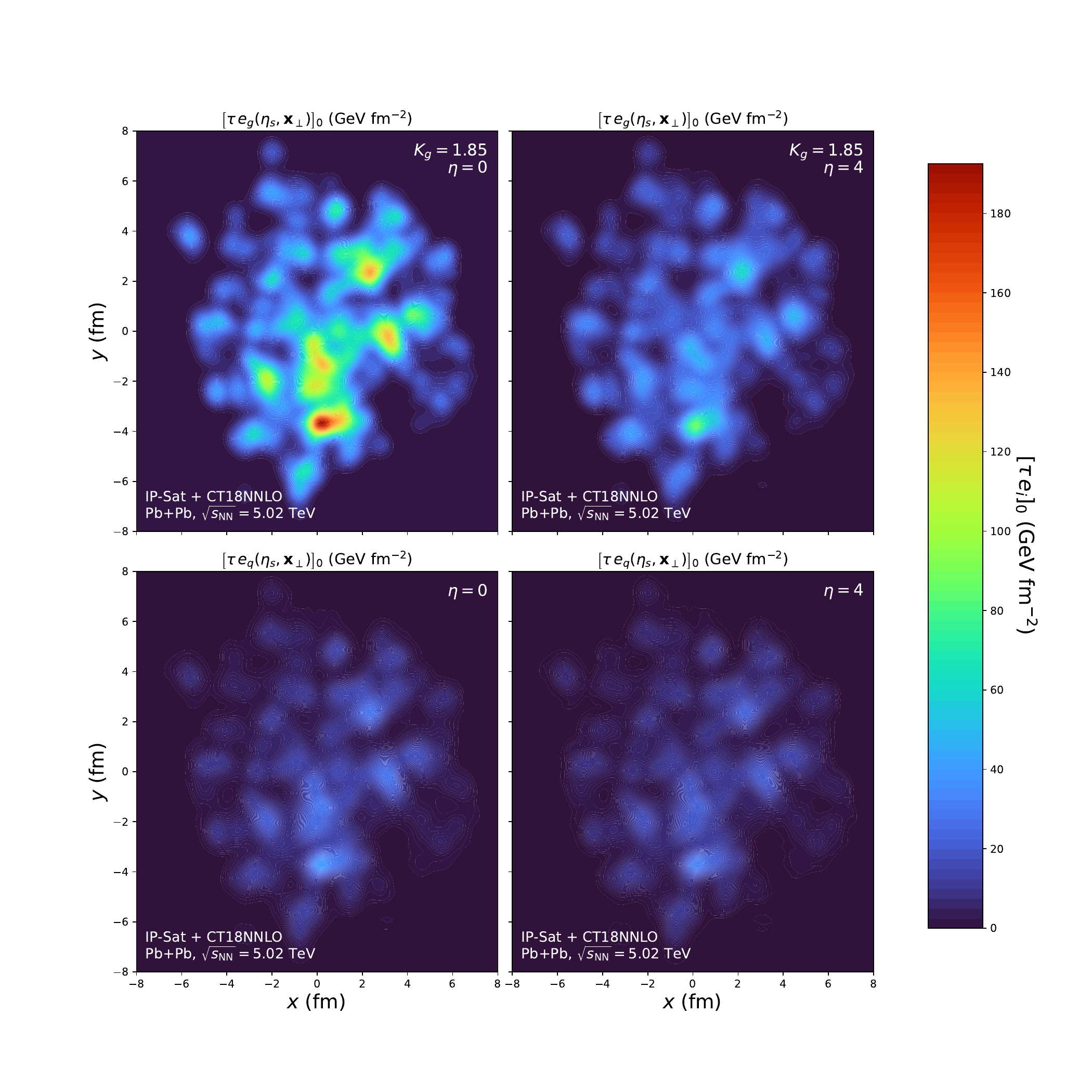}\includegraphics[width=.52\linewidth]{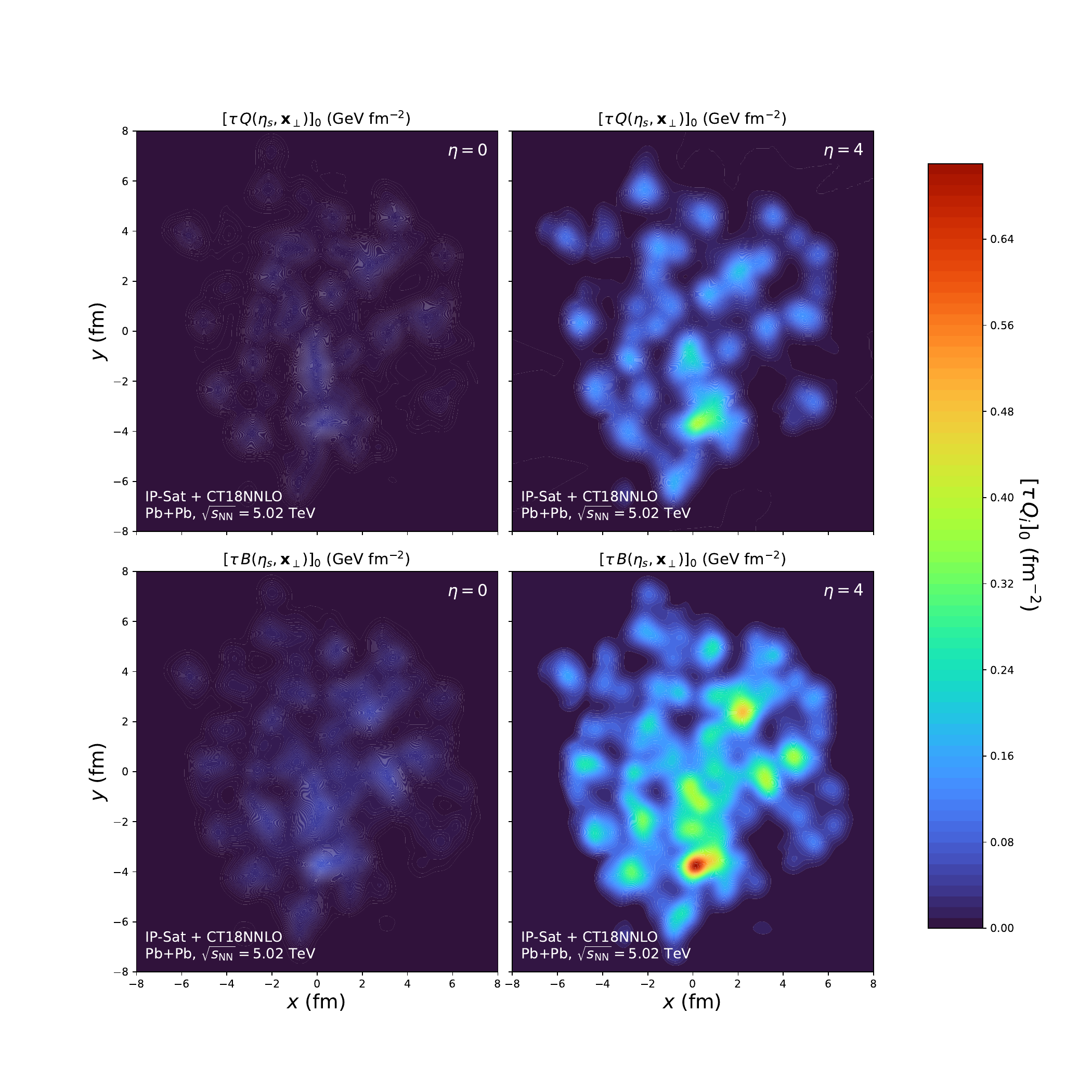}
 \end{center}
  \caption{Energy (\textit{a}) and charge (\textit{b}) profiles for different rapidity windows in a single 5.02 TeV Pb+Pb event in the $0-5\%$ centrality class. The energy profiles are given for gluon energy deposition (\textit{a},\textit{upper pannels}), and quark energy density (\textit{a}, \textit{lower pannels}). The rapidity windows are chosen for $\eta=0 $ (\textit{left pannels}) and $\eta=4$  (\textit{left pannels}). In (\textit{b}) the deposited electric (\textit{b}, \textit{upper pannels}) and baryon (\textit{b}, \textit{lower pannels}) charges are shown for the same rapidity windows.   }
\label{fig:single_event_slices}
\end{figure*}

Now that we have characterized the most important properties our initial state model at mid rapidity, we continue with the exploration of the three dimensional structure of the model. Before we investigate any concrete observables, it proves insightful to illustrate the features of the model at the example of the energy and charge profile in a single event. Different panels in the left part of the figure show the energy deposition due to quarks(bottom) and gluons (top) at mid-rapidity (left) and forward rapidity (right); the right part of the figure shows the corresponding electric (top) and baryon (bottom) charge deposition profiles at mid-rapidity and forward rapidity. Generally, the geometric fluctuations of the energy and charge profiles are dictated by the positions of the nucleons inside the colliding nuclei.
However, due to the different dependencies on the nuclear thickness of the two nuclei (c.f. Sec.~\ref{sec:EbE_MC}), these fluctuations leave a different imprint on the energy and charge profiles at different rapidities. One observes for example, that the relative fluctuations of the energy density profile are larger at mid-rapidity than at forward rapidity. Similarly, the geometric fluctuations of the charge density are also different than those of the energy density, and on an event-by-event basis none of the charge profiles is symmetric w.r.t to the forward and backward rapidity direction. Since all of these features naturally emerge within our framework, it will be interesting to further explore the phenomenological consequences of these findings in the future.

\begin{figure*}[t]
  \includegraphics[scale=0.5]{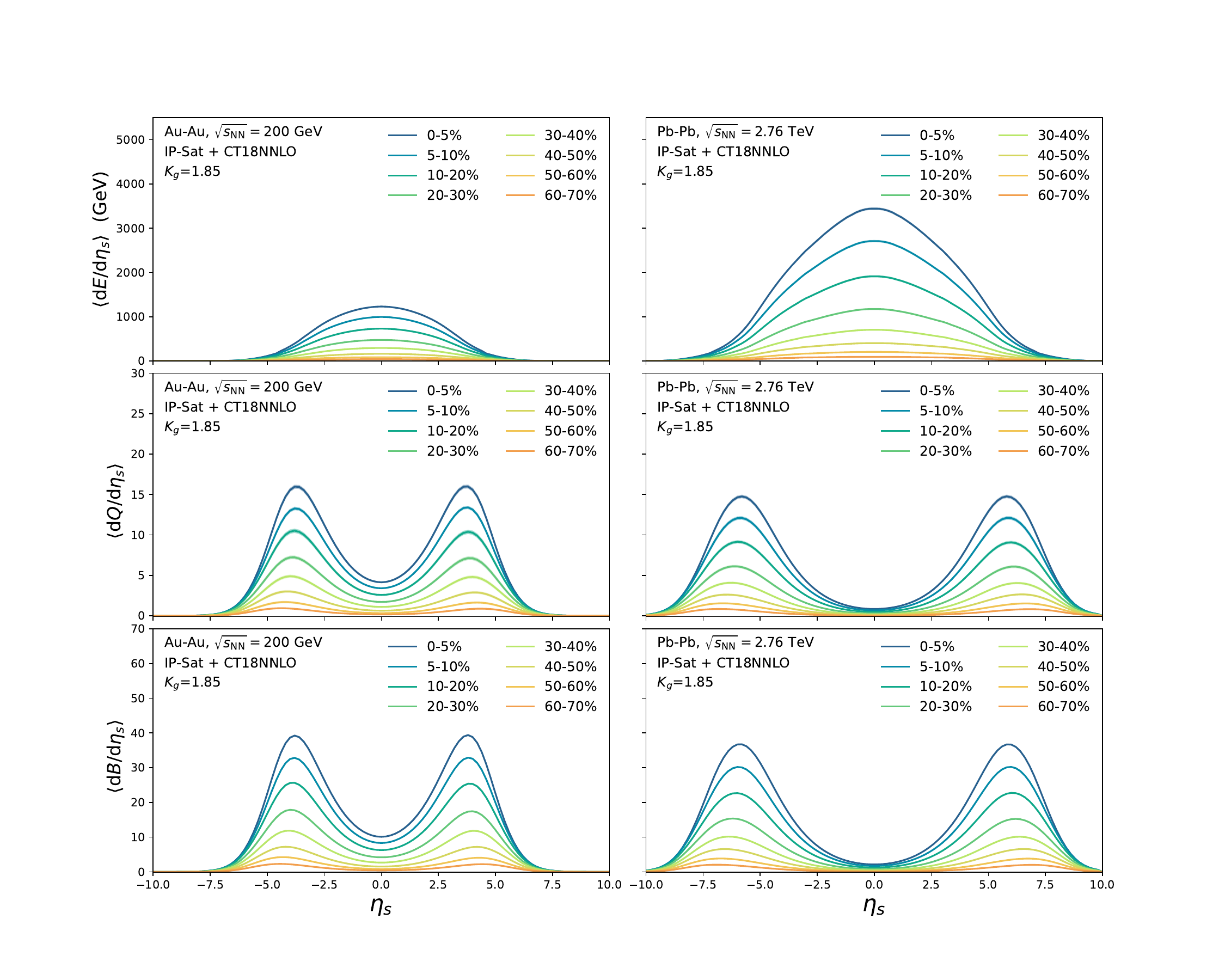}
  \caption{Transverse deposition of energy (\emph{top pannels}), electric charge (\emph{central pannels}) and baryon charge (\emph{lower pannels}). Results are presented for Au-Au collisions at $\sqrt{s_{\rm NN}} =200$ GeV (\emph{left pannels}) and Pb-Pb collisions at $\sqrt{s_{\rm NN}} =200$ GeV (\emph{right pannels}) for different centrality classes. The computation is performed using the IP-Sat model with the CT18NNLO PDF set. }
  \label{fig:dEQBdeta_all_centralities}
\end{figure*}

Next we consider the rapidity dependence of the energy and charge deposition profiles in Fig.~\ref{fig:dEQBdeta_all_centralities}, which we obtain by averaging over several events and integrating over transverse space. Different columns in Fig.~\ref{fig:dEQBdeta_all_centralities} show the results for RHIC and LHC energies, while different curves in each panel show the results for different centralities. While the energy profiles are relative flat around mid-rapdity, they do not exhibit a clear mid-rapidity plateau, which as discussed in Sec.~\ref{sec:anatomy} can primarily be attributed to the rapidity dependence of the gluon energy deposition. Charge deposition profiles for the net-electric and net-baryon charge are peaked at forward and backward rapidities, with the peaks moving out to larger rapidities with increasing center of mass energy. By comparing the profiles at RHIC and LHC energies, one also finds a stronger suppression of net-baryon number and net-electric charge at mid-rapidity for LHC energies, which is consistent with measurements of the hadrochemistry. However, one should caution that over the course of the space-time evolution of a heavy-ion collision the initial charge deposition profiles are subject to diffusion, and can thus not be directly compared with experimental data.

\begin{figure}
  \includegraphics[scale=0.45]{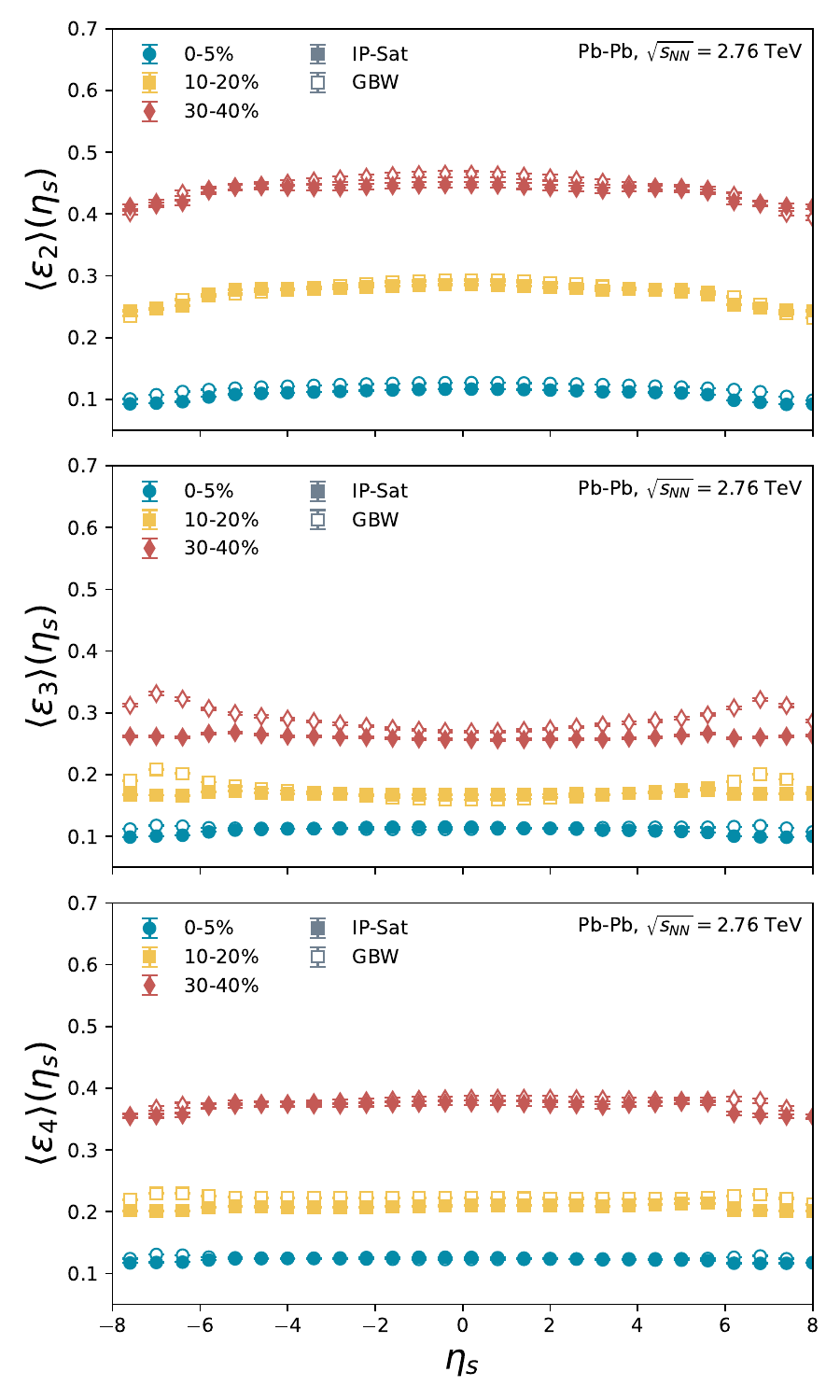}
  \caption{Eccentricities in the IP-Sat (full symbols) and GBW (empty symbols) models as a function of space-time rapidity, $\eta_s$. We show here the $n=2,3,4$ harmonics (in descending order)  for Pb-Pb collisions at $2.76$ TeV and three centrality classes, $0-5\%$, $10-20\%$ and $30-40\%$.}
  \label{fig:eps_rap}
\end{figure}

Despite the fairly pronounced rapidity dependence of the energy density and charge profiles, we find that the rapidity dependence of the spatial eccentricities is rather weak. This can be observed from 
Fig. \ref{fig:eps_rap}, in which the three panels show the rapidity dependence of the event-by-event eccentricities $\varepsilon_2\{2\}$,$\varepsilon_2\{3\}$, and $\varepsilon_2\{4\}$, for three selected centrality classes, $0-5\%$, $10-20\%$ and $30-40\%$. Except at very large rapidities in the forward and backward direction, the eccentricities are almost flat across the entire range of rapidities. Notably, in this regime, the results of the GBW and IP-Sat saturation models can also differ significantly from each other, which can be attributed to the fact that the two models exhibit a rather different behavior at large $x$, which is probed in the far forward and backward regions.

\begin{figure*}
  \includegraphics[scale=0.4]{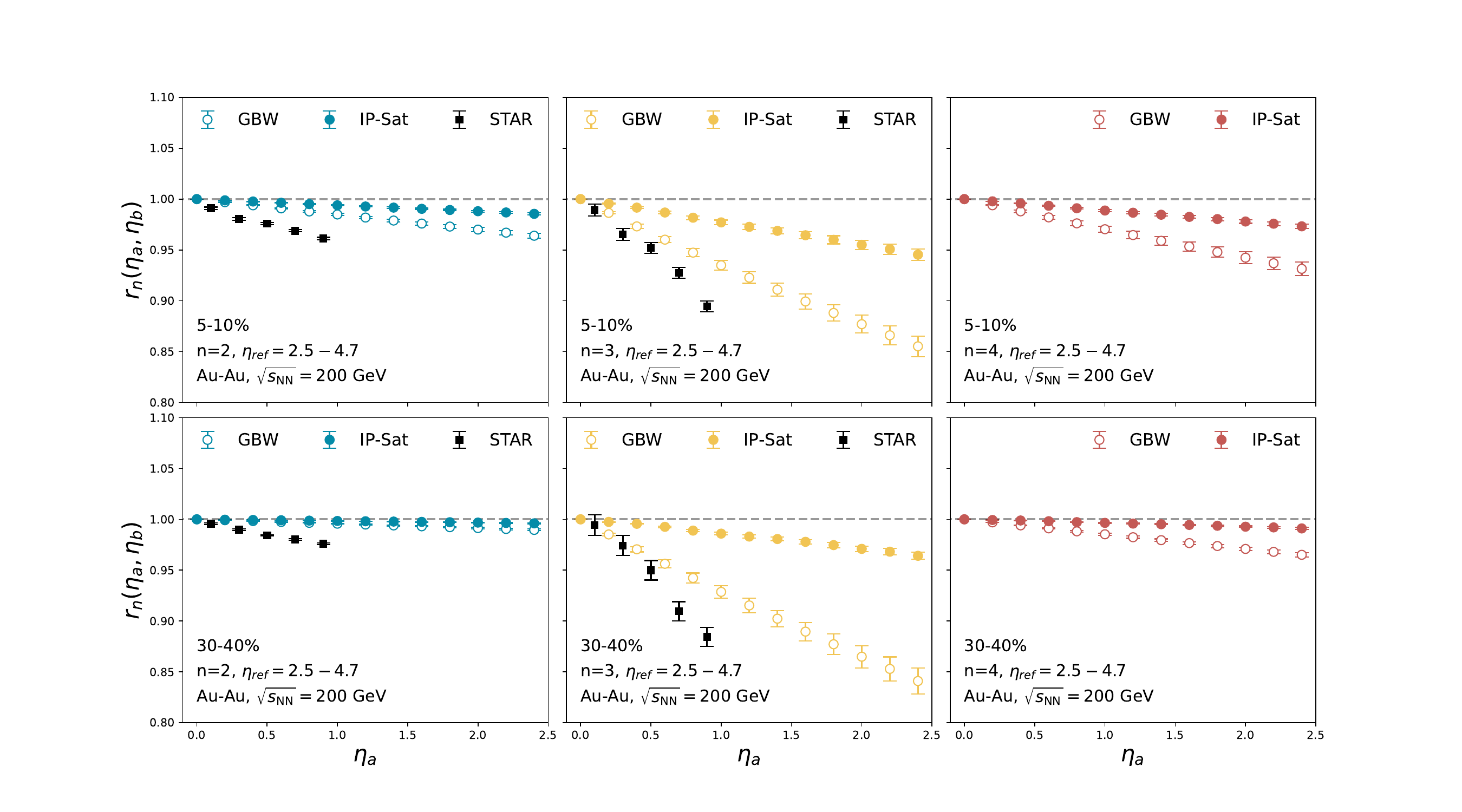}
  \caption{Flow decorrelation of initial spatial eccentricities, $r_n$, in Au-Au collisions $\sqrt{s}=200$~GeV for the GBW and IP-Sat models. In this figure we present the results for the harmonics $n=2,3,4$ (\emph(left), \emph(center) and \emph(right) panels, respectively) and a reference rapidity of $\eta_b=2.5-4.7$ \cite{Nie:2019bgd}}
  \label{fig:flow_dec_RHIC}
\end{figure*}

\begin{figure*}
  \includegraphics[scale=0.4]{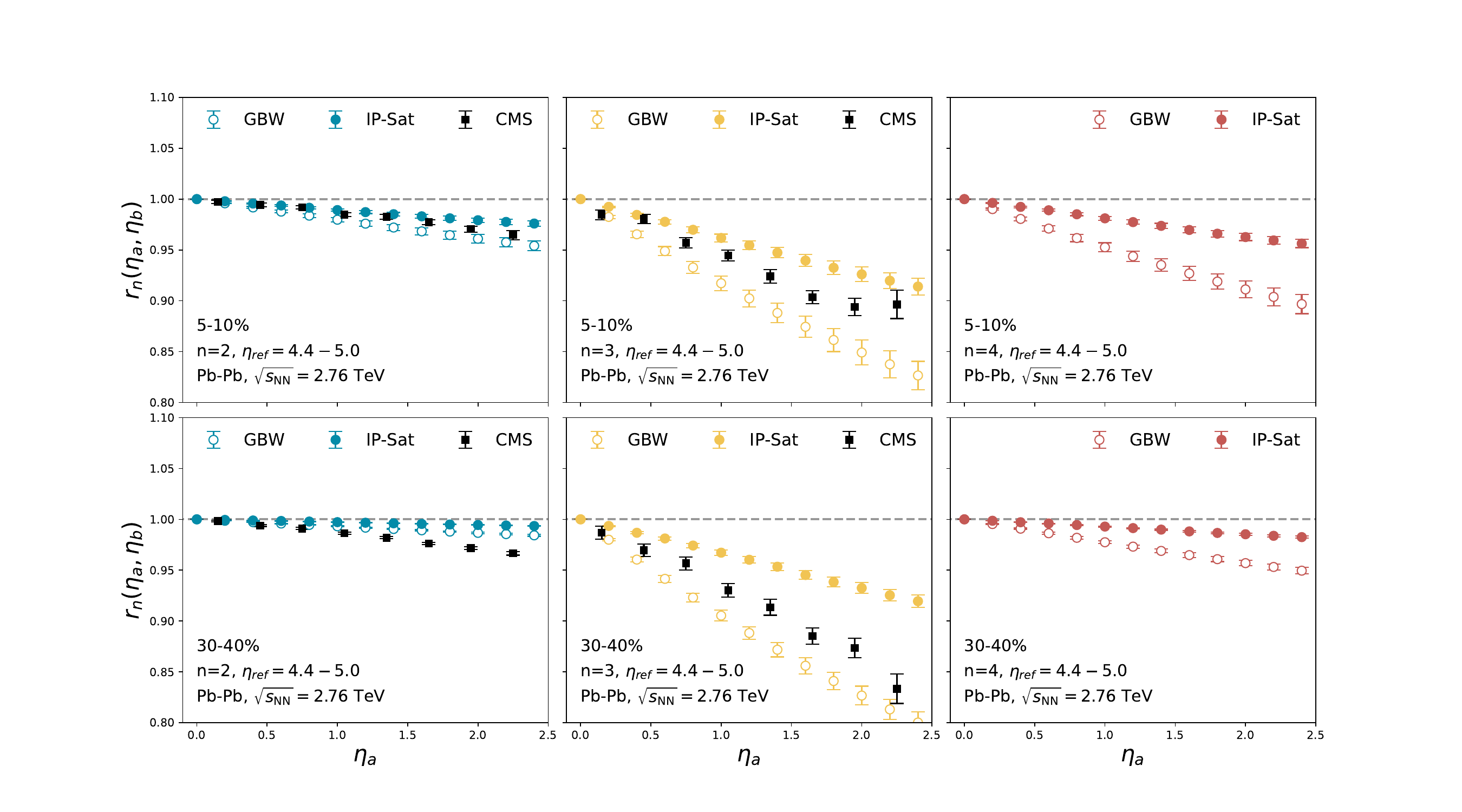}
  \caption{Comparison of the decorrelation of initial spatial eccentricities,$r_n$, of Pb-Pb collisions $\sqrt{s}=2.76$ TeV for the GBW and IP-Sat models. In this figure we present the results for the harmonics $n=2,3,4$ and a reference rapidity of $\eta_b=4.4-5.0$. The black squares correspond to experimental data from the CMS collaboration~\cite{CMS:2015xmx}.}
  \label{fig:flow_dec_CMS}
\end{figure*}

Ultimately, we also investigate the longitudinal decorrelation of the event geometry, which following the experimental procedure~\cite{Nie:2019bgd,CMS:2015xmx} can be quantified by the forward/backward ratio
\cite{Petersen:2011fp,Pang:2018zzo,Pang:2015zrq}
\begin{equation}
    r_n(\eta_a,\eta_b)= \frac{\left\langle\epsilon_n(-\eta_a)\epsilon^*_n(\eta_b)\right\rangle}{\left\langle\epsilon_n(\eta_a)\epsilon^*_n(\eta_b)\right\rangle}
\end{equation}
Our results are compactly summarized in Figs.~\ref{fig:flow_dec_RHIC} and ~\ref{fig:flow_dec_CMS}, where we present the decorrelation ratios $r_2$ (left), $r_3$ (middle) and $r_4$ (right) for $5-10\%$ (top) and $30-40\%$ centrality (bottom) at RHIC and LHC energy. Besides the results from the \Dipper, we also show experimental measurements from STAR~\cite{Nie:2019bgd} (preliminary) and CMS~\cite{CMS:2015xmx} for the ratios $r_{2}$ and $r_{3}$ based on the final state $v_{n}$s. We first note that in comparison to experimental data, our model generally tends to underestimate the de-correlation of the event geometry with varying $\eta_{a}$. One obvious explanation is that the model is missing additional sources of fluctuations; in particular the current model treats the valence content of the nuclei in terms of inclusive (average) PDFs and does not take into account event-by-event fluctuations of the energy-momentum fractions carried by valence partons. Naturally including such fluctuations, e.g. along the lines of~\cite{Shen:2022oyg}, will induce additional sources of geometric fluctuations in the model and thereby increase the slope of the rapidity decorrelation. We also observe large discrepancies for the longitudinal decorrelation between the IP-Sat and GBW model, which we found can be attributed to the rather different large $x$-behavior in these models\footnote{The approach $x\rightarrow 1$ can be quantified by taking the second moment of the dipole function, namely $\langle \kT^2\rangle_i= \int \rmd^2\kT \kT^2\, D_{\rm i}(x,\xT,\qT)/\int \rmd^2\kT \, D_{\rm i}(x,\xT,\qT)$, where the representation is given by $i=(\rm adj, fun)$ for the fundamental case. In the gaussian GBW case, eq.~\eqref{eq:IPSatDip}, this corresponds trivially to $Q^2_{i}\sim (1-x)^\delta$, with $\delta =1$ as defined. Nevertheless, the IP-Sat corresponds to a more nuance case, as the effective extracted $\delta$ presents dependence on the the thickness function, with $\delta\sim 4.6-5.6$. }, which is probed at the reference rapidity $\eta_{b}$. In fact when comparing the results across such large rapidity intervals, the underlying physics changes dramatically, as e.g. in our model the energy densities in the forward regions are no longer dominated by gluon radiation, but rather by quark stopping. Generally speaking, the simultaneous description of the mid-rapidity and fragmentation region provides a challenge for any theoretical description that is not QCD. Even though the choice of a large reference rapidity $\eta_{b}$ certainly has advantages from an experimental point of view, it would be very interesting and highly informative to perform such measurements within a more restricted rapidity range, where theoretical models developed for the physics around mid-rapidity remain applicable.

\section{Conclusions}
\label{sec:conclusions}
We presented a novel approach to calculate energy and charge deposition in high-energy heavy-ion collisions, based on the leading order cross-section calculations in the Color Glass Condensate effective theory of high-energy QCD. Since the underlying quark and gluon distributions can be independently constrained from DIS measurements, the resulting model -- {\Dipper} -- only features a single free parameter $K_{g}$, that is supposed to account for higher order corrections to the leading order perturbative cross-sections. 

Based on this framework, the model makes concrete predictions for the dependence of energy and charge deposition on the nuclear thickness, center of mass energy and space-time rapidity, which naturally emerge from the $x$,$Q^2$ and $T$ dependence of the underlying (nuclear) parton distributions. We therefore believe that the model is not only useful for immediate phenomenological applications, but may also aid the development of parametric models of the longitudinal and transverse structure of the initial state of high-energy heavy-ion collisions. 

Our first comparisons with experimental measurements show promising results, except for the longitudinal decorrelation of the transverse geometry which is underestimated in the current implementation of the model. However, as we discussed this may hint at the necessity to include additional sources of event-by-event fluctuations, in particular of the valence quark distributions, which provides an interesting direction for future investigations. 

Based on our analysis, we also noticed that the physics in the far forward and backward region, which is dominated by nuclear fragmentation, is distinctly different than the physics around mid-rapidity, which is dominated by new particle production, making it hard to describe and interpret measurements that correlate the different rapidity regions. Even though the development of a comprehensive model remains an interesting theoretical challenge, it may also be beneficial to study e.g. de-correlation observables within a more restricted rapidity range, where the underlying physics remains the same.

We finally note that the model, as it is firmly grounded in QCD, is in principle systematically improvable by including higher-order perturbative corrections~\cite{Gelis:2003vh} and sub-leading terms in the eikonal expansion~\cite{Chirilli:2018kkw}. Even though this is a rather challenging task, we believe that this is still a promising avenue, as with the operation of a future Electron-Ion Collider, a more and more precisely knowledge on the underlying (nuclear) parton distributions will emerge.

\section*{Acknowledgements}
OGM would like to thank Travis Dore, Philip Plaschke and Stephan Ochsenfeld for discussions. We would like to thank Bjoern Schenke and Wenbin Zhao for providing the IP-Glasma data shown in this work. Additionally we would like to thank Clemens Werthmann for providing us with the {\trento}  data. 
This work is supported by the Deutsche Forschungsgemeinschaft (DFG, German Research
Foundation) through the CRC-TR 211 ‘Strong-interaction matter under extreme conditions’ — project number 315477589 — TRR 211. OGM and SS acknowledge also support by the German Bundesministerium für Bildung und Forschung (BMBF) through Grant No. 05P21PBCAA.

\appendix
\section{Collinear limit}
\label{app:collinear}
In this section we will show a brief derivation of the collinear limit for the gluon production formula, eq.~\eqref{eq:gluon_prod_w_uGDFs}. This formula has also been derived in the hybrid formalism \cite{Dumitru:2002qt}, and it is valid for gluon forward/backward kinematics, where a large-$x$ gluon scatters of a dense target. For us this means particularly the large $|\eta_s|$ regions. 

We start by integrating the momentum conservation delta function, 
\begin{equation}
\begin{split}
\frac{dN_{g}}{d^2\xT d^2\pT dy}=  &\frac{g^2\,N_c}{4\pi^5 (N_c^2-1)\,\pT^2}  \int \frac{\rmd^2\qT}{(2\pi)^2}\\
&\times~\Phi_1(x_1,\xT,\qT)~\Phi_2(x_2,\xT,\pT-\qT)
\end{split}
\label{eq:gluon_prod_w_uGDFs_delta_int}
\end{equation}

For simplicity we will focus specifically in the forward region, where $\eta_s>0$ and large. In this case, the right moving nucleon has $x_1\sim 1$ while the left moving nucleon possesses the opposite behaviour, $x_2 \ll 1$. Since the characteristic scale, the saturation scale, of these distribution increases with decreasing x, at forward rapidities $\Phi_1$ will peak at small values of $k_\perp$. Since $\Phi_2$ will be dominated by large modes we can take the limit in which $|\pT|>|\qT|$ and expand around it. For the purposes of this work, we can keep only the zeroth order, $\Phi_2(x_2,\xT,\pT-\qT)\rightarrow\Phi_2(x_2,\xT,\pT)$, while higher corrections come as gradients in $\qT$. In this way we get a simplified formula, namely

\begin{equation}
\begin{split}
\frac{dN_{g}}{d^2\xT d^2\pT dy}=  &\frac{1}{(2\pi)^2}  x\,g_A(x_1,\xT,\pT)\,D_{\rm adj}(x_2,\xT,\pT)
\end{split}
\label{eq:collinear }
\end{equation}

where we have defined the nuclear PDF as 
\begin{equation}
\begin{split}
xg_A(x,\xT,\pT^2)=&\frac{N_c^2-1}{16\pi^4 \alpha_s N_c}~\int^{|\qT|<|\pT|}\rmd^2\qT\\
&\,\,\times \qT^2~D_{\rm adj}(x,\xT,\qT)
\end{split}
\label{eq:collinearPDF}
\end{equation}
Notice that our definition of the gluon distribution is defined with dependence to the transverse position in the nucleus, which makes it in fact a parton density. For the case of the total nuclear PDF, we start by integrating with respect to the transverse position $\xT$, 
\begin{equation}
\begin{split}
xg_A(x,\pT^2)=&\frac{N_c^2-1}{16\pi^4 \alpha_s N_c}~
\int\rmd^2\xT\int\rmd^2\qT\\
&\,\,\times \qT^2~D_{\rm adj}(x,\xT,\qT)\,\\
=&\frac{N_c^2-1}{16\pi^4 \alpha_s N_c}~
\int\rmd^2\xT\int\rmd^2\sT\int\rmd^2\qT\,\,\qT^2\\
&\,\,\times e^{\rmi \qT\cdot\sT}~\frac{1}{N_c^2-1} \tr_{\rm adj}[U^{\rm adj}_{\xT+\sT/2} U^{\rm adj,\dagger}_{\xT-\sT/2}]\,,
\end{split}
\label{eq:collinearPDFxT}
\end{equation}
where we have used the definition of the dipole, eq.~\ref{eq:dipole}, in the last line. By defining $\xTp=\xT+\sT/2$ and $\xTm=\xT-\sT/2$, we can transform our formula to 
\begin{equation}
\begin{split}
xg_A(x,\pT^2)=&\frac{1}{16\pi^4 \alpha_s N_c}~
\int\rmd^2\xTp\int\rmd^2\xTm\int\rmd^2\qT\\
&\, e^{\rmi \qT\cdot(\xTp-\xTm)}\,\tr_{\rm adj}[(\partial_{i}U^{\rm adj}_{\xTp})(\partial_i U^{\rm adj,\dagger}_{\xTm})]\,.
\end{split}
\label{eq:collinearPDFxT2}
\end{equation}
In the last line is obtain via partial integration. Using the definition of the Wilson line 
\begin{equation}
U^{\rm adj }_\xT = \exp\left[-\rmi g\int_{-\infty}^{\infty}\rmd x^{+}\,A^-_a(x^+,\xT)\,T^a \right]\,,
 \end{equation}
where $T^a$ are the generators of the $SU(3)$ in the adjoint representation, one can express the  trace in eq.~\eqref{eq:collinearPDFxT2} in the dilute limit to be 
\begin{equation}
\begin{split}
 \tr_{\rm adj}[(\partial_{i}U^{\rm adj}_{\xTp})(\partial_i U^{\rm adj,\dagger}_{\xTm})] = g^2\,N_c\int_{-\infty}^{\infty}\rmd x_1^{+} \\
\times\int_{-\infty}^{\infty}\rmd x_2^{+}\left\langle(\partial_i A^-_{a}(x_1^+,\xTp))(\partial_i A^-_{a}(x_2^+,\xTm)) \right\rangle
\end{split}
\end{equation}
Mixing all together, we get the expression 
\begin{equation}
\begin{split}
xg_A(x,\pT^2)=&\frac{2}{(2\pi)^3}~
\int\rmd^2\qT\,\qT^2 \int\rmd^2\xTp\int\rmd^2\xTm\\
&\times\int_{-\infty}^{\infty}\rmd x_1^{+}
\int_{-\infty}^{\infty}\rmd x_2^{+}\, e^{\rmi \qT\cdot(\xTp-\xTm)}\\ 
&\times \left\langle A^-_{a}(x_1^+,\xTp) A^-_{a}(x_2^+,\xTm) \right\rangle
\end{split}
\label{eq:collinearPDFxT2}
\end{equation}
This result is consistent to the definition of the gluon transverse momentum distribution (TMD) in ref.~\cite{Dumitru:2018kuw}, which can be obtained from the definitions of the TMD found in the literature, see \cite{Petreska:2018cbf,Mulders:2000sh,Marquet:2016cgx,Kotko:2015ura} and references therein.





\bibliographystyle{apsrev4-1}

\bibliography{References}

\end{document}